\documentclass[reprint,amssymb,superscriptaddress]{revtex4-2}

\usepackage{bm}
\usepackage{amsmath}
\numberwithin{equation}{section}
\usepackage[colorlinks=true]{hyperref} \hypersetup{citecolor=red,linkcolor=blue,urlcolor=blue}
\usepackage{hyperref}
\usepackage{dcolumn}

\usepackage{xcolor} \pagecolor{white}
\usepackage{graphicx}

\usepackage[normalem]{ulem}

\newcommand{\respI}[1]{{\color{black}#1}}
\newcommand{\respII}[1]{{\color{black}#1}}

\def\hit{{\rm hit}}
\def\miss{{\rm miss}}
\def\DT{{\rm DT}}
\def\diff{{\rm diff}}

\def\ss{{\rm ss}}

\begin{document}

\title{Optimality of intercellular signaling: direct transport versus diffusion} 
\date{\today}

\author{Hyunjoong Kim}
\affiliation{Center for Mathematical Biology \& Department of Mathematics, University of Pennsylvania, Philadelphia, PA 19104, USA}

\author{Yoichiro Mori}
\affiliation{Center for Mathematical Biology \& Department of Mathematics, University of Pennsylvania, Philadelphia, PA 19104, USA}
\affiliation{Department of Biology, University of Pennsylvania, Philadelphia, PA 19104, USA}

\author{Joshua B. Plotkin}
\affiliation{Center for Mathematical Biology \& Department of Mathematics, University of Pennsylvania, Philadelphia, PA 19104, USA}
\affiliation{Department of Biology, University of Pennsylvania, Philadelphia, PA 19104, USA}

\begin{abstract}
Intercellular signaling has an important role in organism development, but not all communication occurs using the same mechanism. 
Here, we analyze the energy efficiency of intercellular signaling by two canonical mechanisms: diffusion of signaling molecules and direct transport mediated by signaling cellular protrusions.
We show that efficient contact formation for direct transport can be established by an optimal rate of projecting protrusions, which depends on the availability of information about the location of the target cell.
The optimal projection rate also depends on how signaling molecules are transported along the protrusion, in particular the ratio of the energy cost for contact formation and molecule synthesis.
Also, we compare the efficiency of the two signaling mechanisms, under various model parameters. 
We find that direct transport is favored over diffusion when transporting a large amount of signaling molecules.
There is a critical number of signaling molecules at which the efficiency of the two mechanisms are the same.
The critical number is small when the distance between cells is far, which helps explain why protrusion-based mechanisms are observed in long-range cellular communications.
\end{abstract}

\maketitle

\newpage
\section{Introduction}
Intercellular communication is crucial for maintenance and response to the external environment, allowing development, growth, and immunity. 
However, not all biological communication systems transport signals by the same mechanism. One well-known mechanism is simple diffusion, in which signaling molecules are produced by localized source cells and diffuse through extracellular space before degradation \cite{Akiyama2015}. 
An alternative mechanism, called direct transport (DT), involves signaling molecules that are transported along protrusions \cite{Ober2018} such as cytonemes \cite{Kornberg1999,Kornberg2017,Chen2017,Zhang2019}, tunneling nanotubes \cite{Parker2017}, and airinemes \cite{Eom2017,Kim2022}. 
One natural question is under what condition cells should be expected to utilize one or the other mechanism of signaling.

One or the other mechanism may be selectively favored, over evolution, by achieving better performance.
In the scale of communication between two cells,
first passage time of a signaling molecule to the target can be minimized by parallel search with multiple copies (called ``redundancy principle'') \cite{Holcman2019,Lawley2019Impatient}, which characterizes the fertilization process \cite{Yang2016}. This also can be achieved by resetting and repeating the search, which limits the search perimeter \cite{Evans2020ResetReview,Bressloff2020PRE,Bressloff2021SIAP}. 
At a larger spatial scale with multiple cells, a concentration gradient of signaling molecules can be established in a short time  \cite{Kim2018,Kim2018SIAP,Kim2019PRE,Scholpp2020}, which is robust to parameter variation \cite{Kim2018,Kim2018SIAP} and internal noise \cite{Kim2019PRE}, and even precise under a noisy environment \cite{Fancher2020}. 
However, there are only a few direct theoretic analyses comparing these two fundamentally different mechanisms of signaling (direct transport and diffusion), and most such studies focus on the formation of a concentration gradient \cite{Scholpp2020,Fancher2020}.

One crucial aspect of the fitness of cells and organisms is energy efficiency. 
Two significant sources of energy costs are the synthesis of signaling molecules and the polymerization of cellular protrusions. 
Synthesis cost depends on how many diffusive molecules released from the source cell successfully arrive at the target \cite{Stouthamer1973Energyp,Aoyagi1988Energyp}.
DT involves the polymerization cost that is determined by the total number of polymerization events until protrusions make a contact with the target, which is characterized by the total elongation length of protrusions \cite{Kirschner1986,Gallo2020EnergyP}. 
However, the energy cost of intercellular signaling processes has not been considered much in modeling studies, as most such studies focus on the diffusion process, which does not require energy input once it is synthesized.
In contrast to the diffusion model, DT requires energy costs for contact formation. And yet, once established, a protrusion can securely transport signaling molecules. One natural question is which mechanism, under different parameter values, will be more energetically efficient in total.

\begin{figure}[b!]
\includegraphics[width=8.6cm]{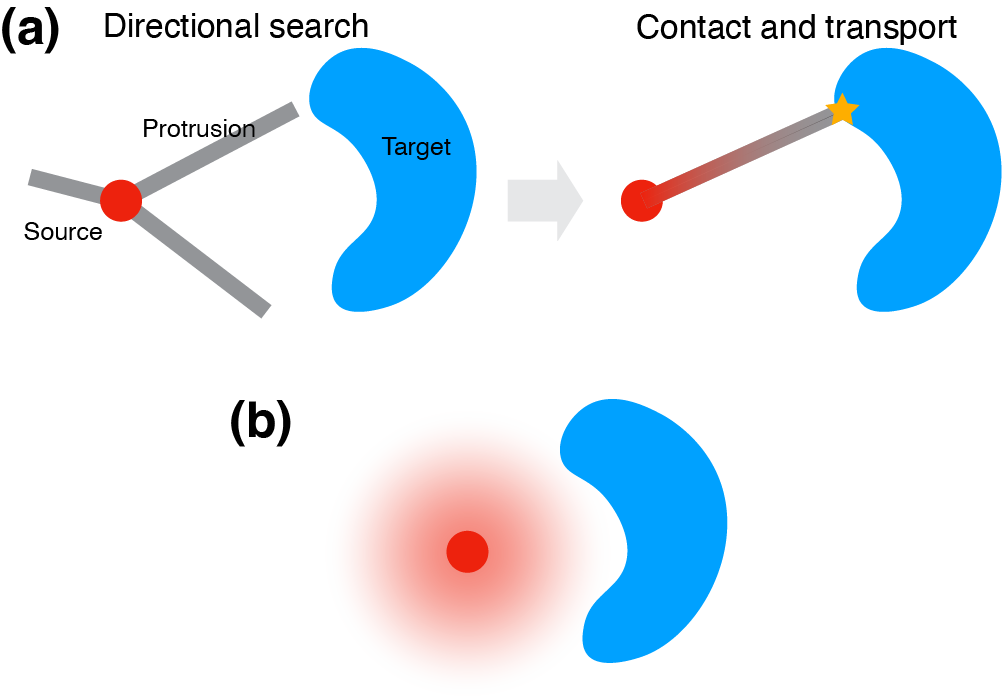}
\caption{\label{fig1} Schematic figure of two intracellular transport mechanisms. (a) Direct transport (DT) mediated by protrusions. (b) Diffusion and degradation of signaling molecules.}
\end{figure}

In this paper, we investigate how a two-cell communication mechanism is optimized by balancing performance (transport time) and energy efficiency. 
We first find the optimal conditions within the direct transport model by minimizing the utility functions associated with the initiation rate of protrusion $\kappa_\DT$. If the protrusion initiation is frequent, then a cell can establish a contact quickly but it may waste energy due to excessive polymerization.
Moreover, the optimal initiation rate depends on the number of signaling particles $V$ required to be transported. This relationship is non-trivial because the effective diffusive transport along finite 1D domain varies with $V$. We show how the relationship changes with the energy consumption rate of polymerization and synthesis of particles.
Additionally, we study how spatial information about the target cell can help form a protrusion contact in a shorter time with less energy cost.
And finally, we study the effect of protrusion length, which, if short, may save polymerization cost but is also less likely to hit the target. Similar optimality considerations for protrusion length versus search time have been studied in other directional search processes with resetting \cite{Bressloff2020PRE,Bressloff2021SIAP}.

Next, we determine which one of the two models (direct transport or diffusion) is optimal for a given condition, by comparing the values of utility functions. One crucial variable for comparison is the number of transporting particles $V$. At small $V$, the diffusion model is generally preferable because it does not require additional energy costs for contact formation. At large $V$, the direct transport model is preferable because the contact formation cost per particle is cheaper. We determine the critical number $V_c$ such that the utility of the two models are the same, which can be a criterion for determining which model is preferable. 

There have been two different approaches to theoretical models of protrusion-based intercellular signaling mechanisms. Early studies \cite{Kim2018,Kim2018SIAP} focused on deterministic continuum models of transporting molecules along with existing signaling protrusion networks. Later studies \cite{Bressloff2020PRE,Bressloff2021SIAP,Kim2019PRE} focused on stochastic search-and-capture models that describe the random search process of signaling protrusions generating signal ``bursts''. Here, our model integrates the two aspects that (i) a source cell first stochastically searches a target cell and establishes a linkage between the source and the target cell, and then (ii) transports molecules along with the linkage, as illustrated in Fig.~\ref{fig1}(a). Our integrated model can quantify the overall signaling time, a sum of stochastic search time and transport time of molecules. In contrast to the previous model, one significant difference in our model is that, instead of a fixed number of nucleation sites for multiple protrusions at a source cell, we assume that cells project protrusions by a Poisson process. Thus, the first passage time (FPT) problem of the multiple protrusions (searchers) now should consider a dependent process. To solve this problem analytically, we approximate the search process by introducing a rare event approximation, which allows taking analytic approaches from \cite{Bressloff2020PRE,Bressloff2021SIAP,Kim2019PRE}. Furthermore, we also consider the total polymerization length until a searcher hits the target (that corresponds to the total length of the searcher’s trajectories), which \respII{is not a linear function} of FPT.

The structure of the paper is as follows. 
In Sect. \ref{Sect2} we introduce a direct transport model that combines the directional search model with resetting \cite{Kim2019PRE,Bressloff2021SIAP} and particle transport model along 1D protrusions \cite{Kim2018,Kim2018SIAP}, as illustrated in Fig.~\ref{fig1}(a).
We introduce two types of idealized targets, a disk and annulus in two dimensions, for analytic simplicity.
We describe a single protrusion search event and then develop a process with multiple search events generated by a Poisson process with the initiation rate $\kappa_\DT$ until contact with the target is formed. We introduce a rare-event approximation of the stochastic contact formation process for analytic simplicity and find relative error bounds. We then quantify the transport time for a required number of diffusive particles along the one-dimensional established protrusion.
In Sect. \ref{Sect3} we introduce the ``mortal'' diffusive model (or diffusive particles with degradation), as depicted in Fig.~\ref{fig1}(b), and we calculate the hitting probability and the transport time similar to the previous section.
In Sect. \ref{Sect4} we define utility functions in terms of performance and energy cost (a cost-benefit ratio and a total energetic cost as a sum of the variables). 
We first investigate the behavior of the cost-benefit ratio for contact formation alone, and we establish the existence of an optimal initiation rate. We then include the process of particle transport along an established protrusion, and we compare the optimal projection rate with and without the particle transport. Finally, we compare the cost-benefit ratio of DT versus mortal diffusion, and we quantify the critical number of signaling molecules that determines which mechanism has more utility. Most of our qualitative results are extended in the case of the total energetic cost.


\section{Direct transport via protrusions} \label{Sect2}
\subsection{Single protrusion event}
\begin{figure}[t!]
\includegraphics[width=8.6cm]{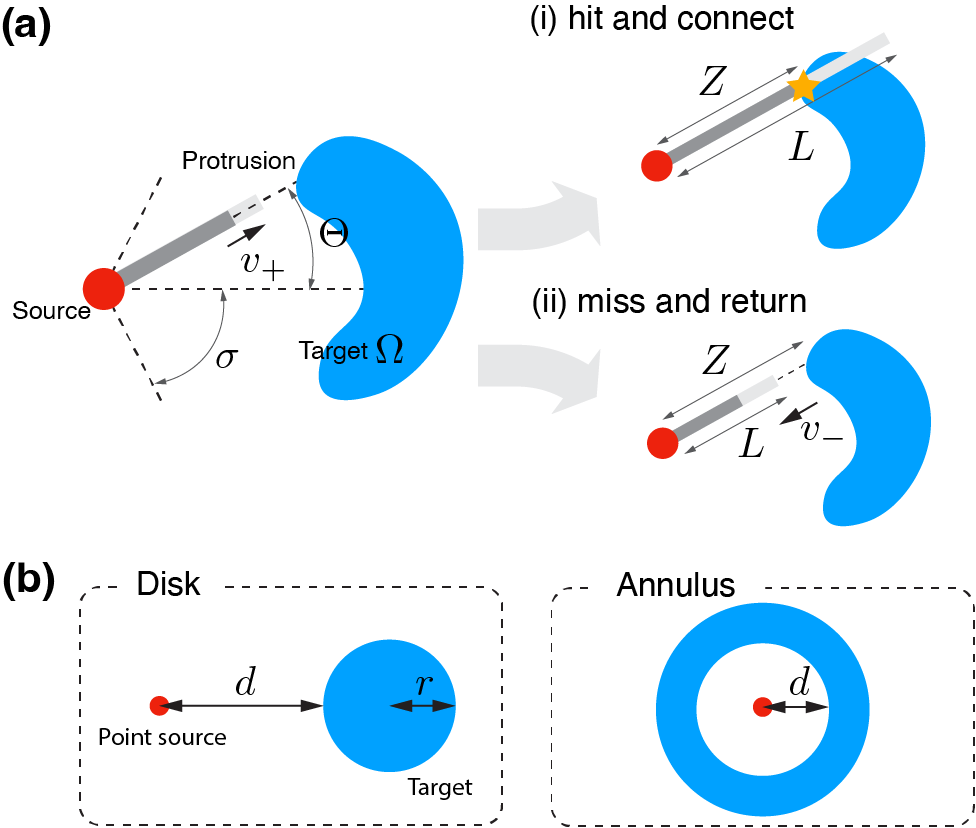}
\caption{\label{fig2} Single protrusion event. (a) Hitting event. A source cell projects a protrusion with random angle $\Theta \in [-\sigma,\sigma]$ at constant speed $v_+$. If random protrusion length $L$ is longer than the effective target distance $Z$, then it hits target $\Omega$. Otherwise, the protrusion grows to $L$ and then returns to the source cell at speed $v_-$. (b) Idealized targets. We consider the idealized shape of cells with a point source and two different types of targets: a disk-shaped target with radius $r$ and minimum distance $d$; a target surrounding the source with distance $d$.}
\end{figure}

Consider a source that extends a protrusion to find target $\Omega$, as illustrated in Fig.~\ref{fig2}(a). 
The protrusion is projected by a random angle $\Theta$ with speed $v_+$. It grows until to a random protrusion length, $L$, or hits the target. If the protrusion fails to hit the target, then it retracts to the source with speed $v_-$.

We first study the contact probability $\rho_\DT$ that the protrusion will hit the target.
The protrusion makes a contact with the target under the following conditions: (i) projection angle $\Theta$ requires to be subtended by the target with respect to the source, and this set of angles is denoted by $\phi(\Omega)$; (ii) the protrusion length $L$ must not be shorter than target distance $\zeta(\Theta)$ at angle $\Theta$. To be well-defined, we set $\zeta(\Theta) = \infty$ if $\Theta \notin \phi(\Omega)$. Introducing the random variable 
\[
    Z = \zeta(\Theta),
\]
which represents the target distance for a projection event,
the hitting probability takes the form of
\begin{equation} \label{rho_DT}
    \rho_\DT = \mathbb{P}[Z \leq L].
\end{equation}

Another important quantity is the actual protrusion length $X$. Since the protrusion stops growing when it hits the target, the actual protrusion length is shorter than the protrusion length $L$. More precisely, the actual protrusion length can be written by
\[
    X = \begin{cases}
        Z, & Z\leq L \\
        L, & Z > L
    \end{cases},
\]
and we denote the protrusion length conditioned on hitting (missing) the target by $X_\hit$ ($X_\miss$).
And so the conditional mean protrusion length satisfies when it hits the target 
\begin{equation}
    \lambda_\hit = \mathbb{E}[X_\hit] = \mathbb{E}[Z|Z\leq L],
\end{equation}
and misses the target
\begin{equation}
    \lambda_\miss = \mathbb{E}[X_\miss] = \mathbb{E}[L|Z > L].
\end{equation}
The mean protrusion length regardless of target contact is
\begin{equation}
    \lambda = \mathbb{E}[X] = \rho_\DT \lambda_\hit + (1-\rho_\DT) \lambda_\miss.
\end{equation}
Moreover, one can also determine the \respI{duration of a single protrusion event} 
in terms of \respI{the actual protrusion length}. Since the protrusion growth and shrinkage speed are assumed to be constant, then the \respI{duration} can be written by
\respI{
\[
    T = \begin{cases}
        Z/v_+, & Z \leq L \\
        L/v_+ + L/v_-, & Z > L
    \end{cases}
\]
and denote the conditional duration when the protrusion hits (misses) the target by $T_\hit$ ($T_\miss$).
}
This yields the conditional \respI{mean duration of a single projection event}
\begin{equation}
    \tau_\hit = \frac{\lambda_\hit}{v_+}, \qquad \tau_\miss = \left(\frac{1}{v_+} + \frac{1}{v_-}\right) \lambda_\miss.
\end{equation}

To compute the statistics of a single protrusion event, we introduce assumptions about the target and the random variables, as depicted in Fig.~\ref{fig2}(b). We consider two types of idealized targets \respI{in 2D}: (i) a disk with minimum distance $d$ and radius $r$ and (ii) an annulus with minimum distance $d$. \respI{The former corresponds to intercellular communication between two distinct cells. The latter corresponds to multicellular communication from a single source cell to multiple target cells. For example, a niche cell controls how quickly neighboring germ cells divide \cite{Inaba2015}.} \respI{We model 2D intercellular communication not only for analytic simplicity but also for representing cell-cell interactions during morphogenesis such as constructing the body axis \cite{Johnston2010} and tracheal organs \cite{Kornberg2017} in \textit{Drosophila}. Though we choose the 2D model, our analysis can be extended to the 3D model.} We assume that protrusion length $L$ follows an exponential distribution with mean $l$, i.e. its distribution takes the form of
\[
    \rho_L(x) = \frac{1}{l} e^{-x/l},
\]
and protrusion angle $\Theta$ follows a uniform distribution with base $2\sigma$ 
\[
    \rho_\Theta(x) = \frac{1}{2\sigma} \mathbb{I}_{[-\sigma,\sigma]}(x).
\]
Here $\mathbb{I}_A(x)$ is the indicator function equal to one if $x \in A$, otherwise zero; and $\Theta = 0$ is set to be the direction of the center of the disk, in the case of the disk-type target.
The computation of the hitting probability and the mean actual protrusion length for the two idealized targets are presented in the Supplementary Material \cite{SM}. 
\respI{In 3D, our analysis can be extended by introducing the spherical coordinate.}

\subsection{Contact formation via multiple protrusion events}
\begin{figure}[t!]
\includegraphics[width=8.6cm]{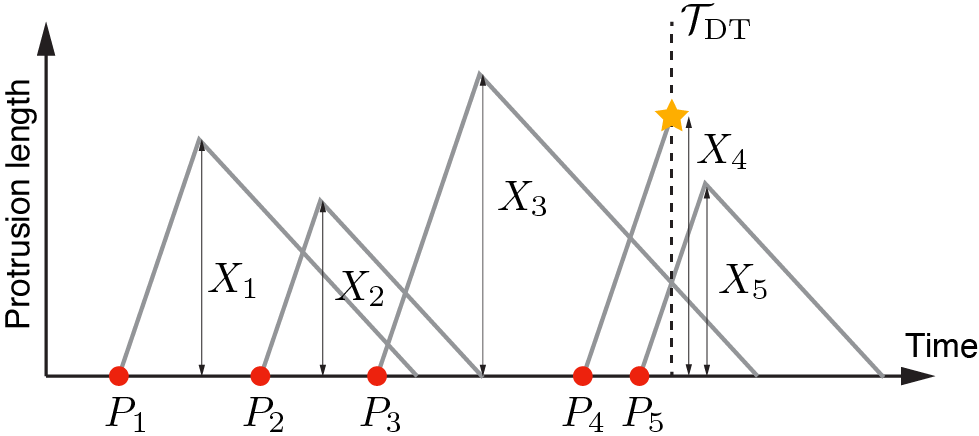}
\caption{\label{fig3} Multiple protrusion events. Source cell projects multiple protrusions at time $P_k$ for $k = 1,2,\cdots$ with exponential inter-projection times $S_k = P_k - P_{k-1}$ at rate $\kappa_\DT$ until the first passage time to the target (FPT) $\mathcal{T}_\DT$. Each protrusion grows by  $X_k$ for $k = 1,2,\cdots$ and the total \respI{polymerization length of protrusions} until the FPT is $\mathcal{X} = \sum_{k : P_k \leq \mathcal{T}_\DT} X_k$.}
\end{figure}

Next, we consider the search process via multiple protrusion events, separated by time intervals that are exponentially distributed with rate $\kappa_\DT$, as depicted in Fig.~\ref{fig3}.
More precisely, let $P_k$ be the $k$th projection time and the corresponding inter-projection times are
\[
    S_k = P_k - P_{k-1}, \qquad S_1 = P_1,
\]
for $k = 1,2,\cdots$. Here $S_k$ are independent and identically distributed exponential times with rate $\kappa_\DT$. The corresponding protrusion event is determined by the pair of the \respI{random} target distance and the protrusion length $(Z_k, L_k)$. 
Let $H_k$ denote the target passage time of the $k$th projected protrusion. If it hits the target, then we have
\[
    H_k = P_k + T_{\hit,k} = P_k + \frac{Z_k}{v_+}.
\]
Otherwise, $H_k = \infty$. \respI{We introduce the set of indices that the protrusion hits the target
\[
    \mathcal{K} = \{k : Z_k \leq L_k \} = \{ k : H_k < \infty\}
\]
}

To determine the overall speed of the target searching process, we are interested in the first passage time (FPT) of protrusions to the target, which takes the form of
\[
    \respI{\mathcal{T}_\DT} = \inf_{k \in \mathcal{K}} \{ H_k\}.
\]
We also compute the total \respI{polymerization length of protrusions} until the FPT, which determines the energy cost for the protrusion polymerization. We introduce the random variable
\[
    \mathcal{N} = \max\{k : P_k \leq \mathcal{T}_\DT\},
\]
which represents the total number of protrusions until the FPT. Then the total polymerization length can be written by
\[
    \respI{\mathcal{X}} = \sum_{k=1}^\mathcal{N} X_k.
\]
\respI{We denote the mean first passage time (MFPT) and the mean total polymerization length by
\[
    \tau = \mathbb{E}[\mathcal{T}_\DT], \quad \xi = \mathbb{E}[\mathcal{X}],
\]
respectively.}

Note that $H_k$ is not necessarily an increasing sequence for $k \in \mathcal{K}$. In other words, even if one protrusion is projected earlier than others that hit the target, it might nonetheless arrive at the target later. Moreover, $\mathcal{X}$ depends on the non-trivial random variable $\mathcal{T}_\DT$. This makes the analysis quite involved, which motivates us to find approximations of these random variables.

\subsection{Rare-event approximation}
\begin{figure}[t]
\includegraphics[width=8.6cm]{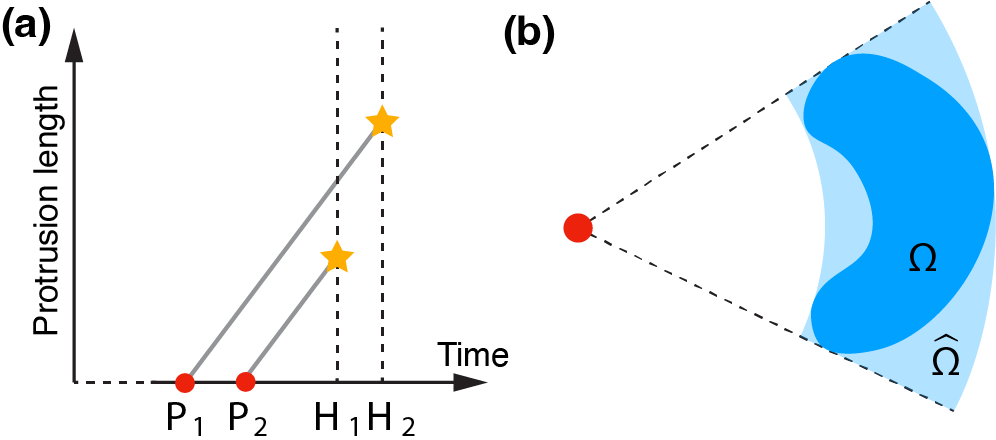}
\caption{\label{fig4} Rare-event approximation. (a) Protrusion departed later might arrive earlier than protrusion departed earlier if $\text{Var}[X_\hit] \neq 0$, and such events are very rare. We approximate the contact formation process by ignoring the rare events and the approximation is exact if $\text{Var}[X_\hit]=0$. (b) Contact formation process to target $\Omega$ is bounded above by the rare-event approximation to target $\Omega$ and below by the approximation to target $\widehat{\Omega}$, which is the minimal ``polar rectangle'' containing $\Omega$.}
\end{figure}

We approximate the important random variables $\mathcal{T}_\DT$ and $\mathcal{X}$ by the rare-event approximation. 
In general, $H_k$ is not an increasing sequence for $k \in \mathcal{K}$, but it is very unlikely that the protrusion projected later hits the target earlier than the protrusion projected earlier, as depicted in Fig.~\ref{fig4}(a). 
Thus, we make the approximation that $H_k$ is increasing. \respI{That is, the rare-event approximation of the first passage time occurred at the first protrusion heading to the target
\begin{equation}
	\widetilde{\mathcal{T}}_\DT = H_{\mathcal{K}_0},
\end{equation}
where $\mathcal{K}_0 = \min\mathcal{K}$. Under the assumption, the source cell projects approximately $\mathcal{K}_0$ protrusions to generate the first protrusions hitting the target. During that protrusion grows to the target ($T_{\hit,\mathcal{K}_0}$), the source cell still generates the protrusions with rate $\kappa_\DT$. Thus, the rare-event approximation of the total number of protrusions can be written by
\begin{equation}
    \widetilde{\mathcal{N}} = \mathcal{K}_0 + \mathcal{N}_0(T_{\hit,\mathcal{K}_0}).
\end{equation}
Here $\mathcal{N}_0(t)$ is the number of protrusions over time $t$ \respI{with rate $\kappa_\DT$}.  Therefore, the rare-event approximation of the total polymerization length is
\begin{equation} \label{tildeL}
    \widetilde{\mathcal{X}} = \sum_{k=1}^{\widetilde{\mathcal{N}}} X_k.
\end{equation}
}

Note that this approximation is exact when the actual protrusion length for hitting the target has zero variance, \respI{$\text{Var}[X_\hit] = 0$}, in which case the rare event we neglect has zero probability. One example is that the source is surrounded by the target at the same distance. In general, the rare-event approximation overestimates the FPT and so does the total polymerization length
\begin{equation} \label{ub}
    \mathcal{T}_\DT \leq \widetilde{\mathcal{T}}_\DT, \qquad \mathcal{X} \leq \widetilde{\mathcal{X}},
\end{equation}
due to the way of approximation. The important advantage of this rare-event approximation is that it provides analytical tractability. In this section, we calculate the first moment of the rare-event approximation of FPT and the \respI{total polymerization length}. Then we perform an error estimation of the rare-event approximation.

We first calculate the rare-event approximation of the MFPT by applying the total expectation theorem \cite{Kim2019PRE}. 
Conditioning \respI{$\widetilde{\mathcal{T}}_\DT$} on $\mathcal{K}_0 = j$ gives
\begin{eqnarray} 
	\widetilde{\tau}_{j} &=& \mathbb{E}[\widetilde{\mathcal{T}}_\DT|\mathcal{K}_0 = j] = \mathbb{E}\left[\sum_{k=1}^j S_k + T_{\hit,j}\right] \nonumber \\
    &=& \frac{j}{\kappa_\DT} + \tau_\hit, \label{tau_DT1}
\end{eqnarray}
according that $T_{\hit,j}$ is independent of $j$.
Using the total expectation theorem yields
\begin{eqnarray*} 
	\widetilde{\tau} &=& \mathbb{E}\Big[\mathbb{E}[\widetilde{\mathcal{T}}_\DT|\mathcal{K}_0] \Big] = \sum_{j=1}^\infty \widetilde{\tau}_{j} \mathbb{P}[\mathcal{K}_0 = j] \nonumber \\
	&=& \sum_{j=1}^\infty \widetilde{\tau}_{j} \rho_{\DT} (1-\rho_{\DT})^{j-1}.
\end{eqnarray*}
Substituting Eq.~(\ref{tau_DT1}) we finally have the rare-event approximation of the MFPT
\begin{equation} \label{t_DT_approx}
	\widetilde{\tau} = \frac{1}{\kappa_{\DT} \rho_{\DT}} + \tau_{\hit}.
\end{equation}
This can be interpreted as the sum of the first projection time of which protrusion hits the target ($\kappa_{\DT}^{-1} \rho_{\DT}^{-1}$) and the protrusion travel time to the target ($\tau_{\hit}$).

We calculate the rare-event approximation of the mean total number of protrusions.
\respII{$\mathcal{N}_0(t)$ is a Poisson point process with rate $\kappa_\DT$, that is, the number of events in any interval of length $t$ is a Poisson random variable with parameter (or mean) $\kappa_\DT t$
\begin{equation}
	\mathbb{P}[\mathcal{N}_0(t) = k] = \frac{(\kappa_\DT t)^k}{k!}e^{-\kappa_\DT t}.
\end{equation}
This property implies that
\begin{equation} \label{nu0}
	n_0(t) = \mathbb{E}[\mathcal{N}_0(t)] = \kappa_\DT t.
\end{equation}
Therefore, we have
\begin{eqnarray}
	\mathbb{E}[\widetilde{\mathcal{N}}] &= \mathbb{E}[\mathcal{K}_0] + \kappa_\DT \mathbb{E}[T_{\hit,\mathcal{K}_0}] \nonumber \\
	&= \frac{1}{\rho_\DT} + \kappa_\DT \tau_\hit = \kappa_\DT \widetilde{\tau},
\end{eqnarray}
which can be interpreted as the total number of protrusions until the FPT.}

We next calculate the rare-event approximation of the mean total polymerization length. 
Eq. (\ref{tildeL}) can be written by
\begin{eqnarray}
    \widetilde{\mathcal{X}} &=& \sum_{k=1}^{\mathcal{K}_0} X_k + \sum_{k=\mathcal{K}_0 + 1}^{\mathcal{K}_0 + \mathcal{N}_0(T_{\hit,\mathcal{K}_0})} X_k,
\end{eqnarray}
where $X_k = X_{\miss,k}$ if $k < \mathcal{K}_0$ and $X_{\mathcal{K}_0} = X_{\hit,\mathcal{K}_0}$.
Since $\mathcal{K}_0$ and $X_{\hit,\mathcal{K}_0}$ are independent we condition the expectation by setting $\mathcal{K}_0 = j$ and $X_{\hit,\mathcal{K}_0} = x$
\begin{eqnarray*}
    \widetilde{\xi}_{j}(x) &=& \mathbb{E}[\widetilde{\mathcal{X}}|\mathcal{K}_0 = j, ~X_{\hit,\mathcal{K}_0} = x] \nonumber \\
    &=& (j-1) \lambda_\miss + x + \mathbb{E}\left[\sum_{k=1}^{\mathcal{N}_0(x/v_+)}X_k\right].
\end{eqnarray*}
Using the independence of $\mathcal{N}_0(x)$ and $X_k$ for $k > j$, we have
\begin{eqnarray}
    \mathbb{E}\left[\sum_{k=1}^{\mathcal{N}_0(x/v_+)}X_k\right] &=& \mathbb{E}[\mathcal{N}_0(x/v_+)] \mathbb{E}[X_k] \nonumber \\
    &=& \frac{\kappa_\DT x \lambda}{v_+}, \label{sumX}
\end{eqnarray}
by substituting Eq.~(\ref{nu0}). Thus, the unconditional expectation satisfies
\begin{eqnarray}
	\widetilde{\xi} &=& \mathbb{E}\left[\mathbb{E}[\widetilde{\mathcal{X}}|\mathcal{K}_0, ~X_{\hit,\mathcal{K}_0}]\right] \nonumber \\
	&=& \sum_{j=1}^\infty  \rho_{\DT} ( 1- \rho_{\DT})^{j-1}\widetilde{\mathcal{\xi}}_{j}(\lambda_\hit)  \nonumber \\
	&=& (\rho_\DT^{-1} - 1) \lambda_\miss + \lambda_\hit + \kappa_\DT \tau_\hit \lambda,
\end{eqnarray}
which can be simplified as
\begin{equation} \label{lambda_tau}
    \widetilde{\xi} = \kappa_\DT \widetilde{\tau} \lambda.
\end{equation}
\respI{This implies that the source cell projects protrusions $\kappa_\DT \tilde{\tau}$ times with average length $\lambda$ to hit the target cell.}

\begin{figure}[t!]
\includegraphics[width=8.6cm]{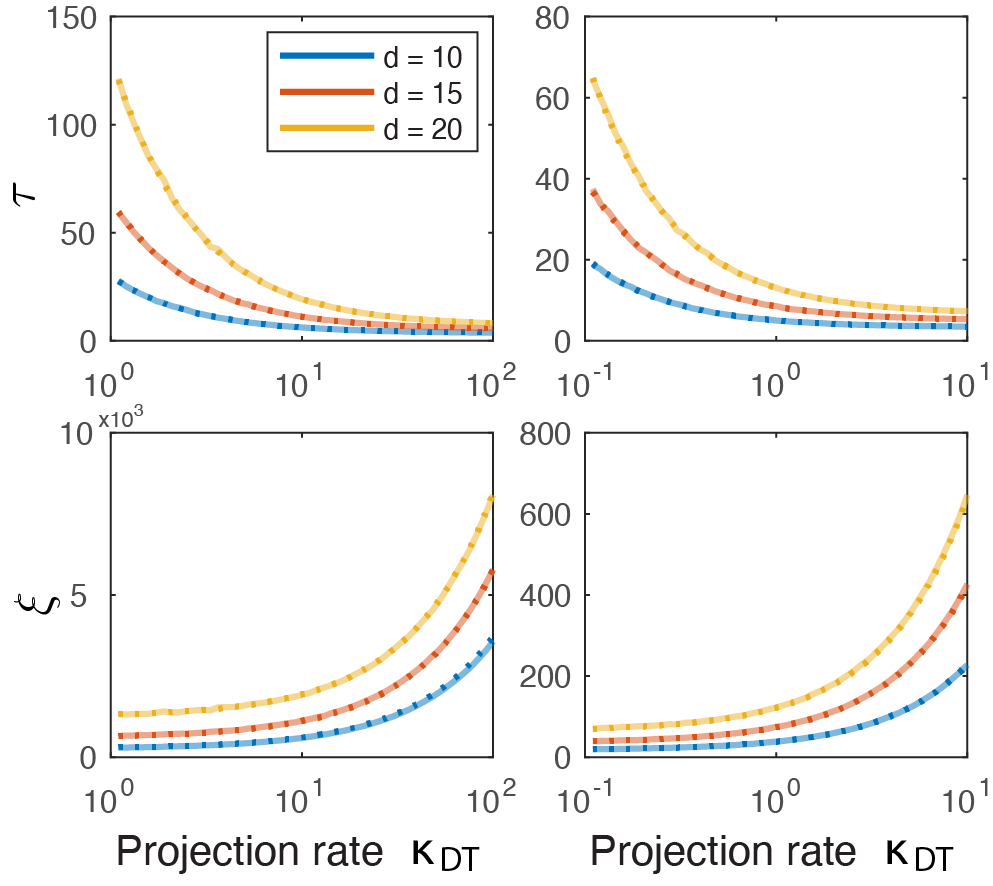}
\caption{\label{fig5} Comparison between the exact (\textit{solid curve}) and the rare-event approximation (\textit{dotted curve}) of the mean first passage time (MFPT) $\tau$ and the mean total polymerization length $\xi$ with various minimum distance $d$ $\mu$m in case of the disk-shaped target (\textit{left panel}) and the annulus-shaped target (\textit{right panel}). Parameters are as follows: $r = 5$ $\mu$m, $l = 10$ $\mu$m, $\sigma = \pi$, $v_+ = 8.5$ $\mu$m/min, and $v_- = 3$ $\mu$m/min.}
\end{figure}
Fig.~\ref{fig5} shows that the rare approximation mean has good agreement with the MFPT and the mean polymerization length estimated by the Monte Carlo simulations. Moreover, direct contact can be formed in a shorter time by generating more protrusions as the projection rate increases. However, the asymptotic MFPT never goes to zero, instead, it converges to
\begin{equation}
    \lim_{\kappa_\DT \to \infty} \tau(\kappa_\DT) = \frac{d}{v_+},
\end{equation}
which is the traveling time along the minimum distance (geodesic), as suggested in \cite{Lawley2021PDMP,Lawley2020GeoDiff}.

One natural question is how close the rare-event approximation is to the exact process. We address this issue by determining an error bound \respI{with a minimal ``polar rectangle'' $\widehat{\Omega}$ containing target $\Omega$, as illustrated in Fig. \ref{fig4}(b). In particular, we proceed with the error analysis for a target disk.} Consider a target disk with radius $r$ and minimum distance $d$ from the source. Then the actual target distance of $\widehat{\Omega}$ is shorter than $\Omega$ for a given angle
\[
    \zeta(\Theta;\Omega) \geq \zeta(\Theta;\widehat{\Omega}) = d, 
\]
\respI{for angle $\Theta \in \phi(\Omega)$. For a minimal polar rectangle, it satisfies 
\[
    \phi(\Omega) = \phi(\widehat{\Omega}).
\]
In other words, the minimal polar rectangle has a shorter distance in the set of subtended angles. This implies that}
the hitting probability of a single protrusion event to $\widehat{\Omega}$ is larger than $\Omega$ and shorter than the conditional hitting time
\begin{equation}
    \rho_\DT(\Omega) \leq \rho_\DT(\widehat{\Omega}), \qquad \tau_\hit(\Omega) \geq \tau_\hit(\widehat{\Omega}),
\end{equation}
which also guarantees that
\begin{equation} \label{lb}
    \mathcal{T}_\DT(\widehat{\Omega}) \leq \mathcal{T}(\Omega),
\end{equation}
for the same sequence of random pairs $(\Theta_k,L_k)$ for $k = 1,2,\cdots$. Since $T_\hit(\widehat{\Omega}) = d$ and $\text{Var}[T_\hit(\widehat{\Omega})] = 0$, we have
\begin{equation}
    \tau(\widehat{\Omega}) = \widetilde{\tau}(\widehat{\Omega}).
\end{equation}
Together with Eqs. (\ref{ub}) and (\ref{lb}), we have the error bound of the rare-event approximation
\begin{equation}
    |\widetilde{\tau}(\Omega) - \tau(\Omega)| \leq \delta [\widetilde{\tau}],
\end{equation}
where $\delta[f] = f(\Omega) - f(\widehat{\Omega})$ and
\begin{equation}
    \delta[\widetilde{\tau}] = \frac{\delta[\rho_\DT^{-1}]}{\kappa_\DT} + \frac{\delta[\lambda_\hit]}{v_+}
\end{equation}
in accordance with Eq.~(\ref{t_DT_approx}).
Moreover, the relative difference satisfies
\begin{eqnarray}
    \frac{\delta[\widetilde{\tau}]}{\widetilde{\tau}(\Omega)} &\leq& \frac{\delta[\rho_\DT^{-1}]}{\rho_\DT^{-1}(\Omega)} + \frac{\delta[\lambda_\hit]}{\lambda_\hit(\Omega)} \nonumber \\
    &\leq& 1 - e^{-r/l} + \frac{r}{r+d}, \label{rel_err}
\end{eqnarray}
in terms of the model parameters (see \cite{SM} for details). 
This inequality implies that the relative error is smaller if the target is smaller and farther from the source. That is, the rare-event approximation is accurate if it is hard to hit the target in a single protrusion. Similarly, one can derive the error bound for the mean total polymerization length
\begin{equation}
    |\widetilde{\xi}(\Omega) - \xi(\Omega)| \leq \delta[\widetilde{\xi}] = \kappa_\DT \lambda \delta[\widetilde{\tau}],
\end{equation}
according to Eq. (\ref{lambda_tau}). This has the same relative bound in Eq.~(\ref{rel_err}).

\subsection{Particle transport along protrusion} \label{sect2D}
\begin{figure}[t!]
\includegraphics[width=8.6cm]{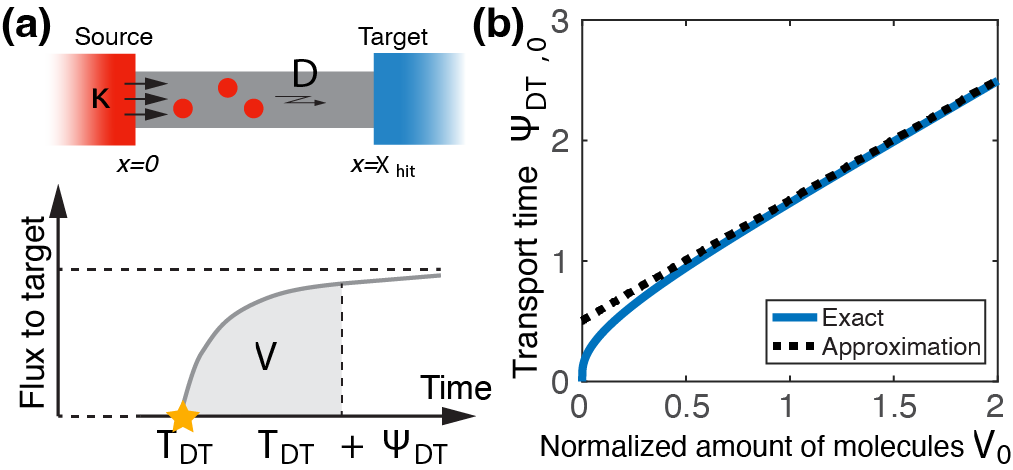}
\caption{\label{fig6} Molecule transport time of the direct transport model. (a) Illustration of the particle transport time. Once establishing a contact at time $\mathcal{T}_\DT$ with length $X_\hit$, diffusive signaling molecules are released at the source end with rate $\kappa$ and absorbed at the target end. It takes additional time $\Psi_\DT$ to transport the required amount of particles $V$. (b) Plot of non-dimensional particle transport time \respI{$\Psi_{\DT,0} =\Psi_\DT D/X_\hit^2$} as a function of non-dimensional particle number \respI{$V_0 = VD/(\kappa X_\hit^2)$}, together with the asymptotic approximation \respI{$\overline{\Psi}_{\DT,0} = \overline{\Psi}_{\DT} D/X_\hit^2$}.}
\end{figure}

Once a protrusion makes a contact with length $X_\hit$, we assume that the source begins to produce signaling molecules with rate $\kappa$ \respI{and the target cell absorbs the molecules}, as depicted in Fig.~\ref{fig6}(a). 
We assume that the molecules diffuse along the protrusion, though transport of proteins in cellular protrusions can also occur by active transport \cite{Chen2017,Cevik2013}.
\respII{By assuming diffusion along the protrusion we can directly compare the efficiency of the same basic transport mode, with the same diffusion constant, under different geometries. The mechanism of direct transport allows molecules to diffuse along a thin ``pipeline'' between cells (a 1D domain), whereas diffusive particles are delivered through the space without geometric restriction (a 2D domain).}

We now determine the transport time $\psi_\DT$ for delivering a required number of signaling molecules, $V$. 
More precisely, let $u(x,t)$ denote the molecule concentration along the protrusion for $0<x<X_\hit$, which evolves according to
\begin{equation} \label{DTdiff}
	\frac{\partial u}{\partial t} = D \frac{\partial^2 u}{\partial x^2},
\end{equation}
where $D$ is the diffusion coefficient along a protrusion. Here we assume that the molecules are not degraded while they move within the protrusion, because diffusion along the protrusion is relatively stable compared to the outer environment. \respII{Thus, we can consider the protrusion as a ``secure pipeline'' along which signalling molecules diffuse.} This equation is supplemented by the boundary conditions
\begin{equation} \label{DTdiff_bc}
	-D \frac{\partial u(0,t)}{\partial x} = \kappa, \qquad u(X_\hit,t) = 0.
\end{equation}
Then the flux at position $x$ takes the form
\[
    J_\DT(x,t) = - D \frac{\partial u(x,t)}{\partial x}.
\]
\respI{Note that the concentration $u$ and the flux $J_\DT$ have the unit of inverse length and inverse time, respectively.}
The transport time $\Psi_\DT(V;X_\hit)$ for given protrusion length $X_\hit$ and required number of particles $V$ satisfies the following integral equation
\begin{equation} \label{int_eq}
    \int_0^{\Psi_\DT} J_\DT(X_\hit,t')dt' = V,
\end{equation}
which yields the mean particle transport time depending on the protrusion length
\[
    \psi_\DT (V) = \mathbb{E}[\Psi_\DT(V;X_\hit)].
\]

Numerically computing the mean particle transport time can be complex because $X_\hit$ is implicitly involved in the expression. We address this issue by introducing the non-dimensional variables $x_0 = x/X_\hit$, $t_0 = tD/X_\hit^2$, $V_0 = VD/(\kappa X_\hit^2)$, and $J_{\DT,0} = J_\DT/\kappa$. The integral equation (\ref{int_eq}) can then be written by
\begin{equation}
    \mathcal{V}_\DT(\Psi_{\DT, 0}) := \int_0^{\Psi_{\DT, 0}} J_{\DT,0}(t_0') dt_0' = V_0,
\end{equation}
where $\mathcal{V}_\DT(\Psi_{\DT, 0})$ represents the non-dimensional transported molecules to the target over non-dimensional time $\Psi_{\DT, 0}$. Taking the Laplace transformation
\[
    \mathcal{L}[f](s) = \int_0^\infty f(t) e^{-st} dt,
\]
of the non-dimensional number of transported particles yields
\begin{equation}
    \mathcal{L}[\mathcal{V}_\DT](s) = \frac{1}{s}\mathcal{L}[J_{\DT,0}](s),
\end{equation}
where
\begin{equation}
    \mathcal{L}[J_{\DT,0}](s) = \frac{1}{s \cosh(\sqrt{s})},
\end{equation}
which does not depends on any model parameters. This also gives the asymptotic flux
\begin{eqnarray}
    J_{\DT,0}^\ss &=& \lim_{t_0\to \infty} J_{\DT,0}(t_0)  \nonumber \\
    &=& \lim_{s \to 0} s \mathcal{L}[J_0](s) = 1.
\end{eqnarray}
One can numerically evaluate $\mathcal{V}_\DT(t)$ by taking the numerical inverse transformation on $\mathcal{L}[\mathcal{V}_\DT](s)$ \cite{Kuhlman2013Laplace}, as shown in Fig.~\ref{fig6}(b). Since $\mathcal{V}_\DT(t_0)$ is monotonically increasing, we deduce
\[
    \Psi_{\DT, 0} = \mathcal{V}_\DT^{-1} (V_0),
\]
which is equivalent to
\begin{equation}
    \Psi_\DT = \frac{X_\hit^2}{D} \mathcal{V}_\DT^{-1}\left(\frac{DV}{\kappa X_\hit^2}\right).
\end{equation}
Therefore, the mean transport time is also can be written by
\begin{equation}
    \psi_\DT(V) = \mathbb{E}\left[\frac{X_\hit^2}{D} \mathcal{V}_\DT^{-1} \left(\frac{DV}{\kappa X_\hit^2}\right)\right].
\end{equation}

When the number of particles $V$ required to transport is large, we can approximate the mean particle transport time by using the fact that the flux converges to a constant. We write the implicit equation as 
\begin{equation} \label{LNA_DT}
    \int_0^{\Psi_{\DT,0}} J_{\DT,0}(t_0') - J_{\DT,0}^\ss dt_0' + J_{\DT,0}^\ss \Psi_{\DT,0} = V_0,
\end{equation}
which can be approximated by
\begin{equation} \label{nondim_int_eq}
    \int_0^\infty J_{\DT,0}(t_0') - J_{\DT,0}^\ss dt_0' + J_{\DT,0}^\ss \overline{\Psi}_{\DT,0} = V_0,
\end{equation}
when $V_0 \gg 1$, that is, the required number of molecules is sufficiently larger than the released molecules over the time interval that the diffusive particle travels the protrusion ($\kappa X_\hit^2/D \ll V$). Taking the Laplace transformation, the first integral term reduces to
\begin{eqnarray}
    \int_0^\infty J_{\DT,0}(t_0') - J_{\DT,0}^\ss dt_0' &=& \lim_{s \to 0} \mathcal{L}[J_0-J_{\DT,0}^\ss](s) \nonumber \\
    &=& \lim_{s \to 0} \frac{1}{s\cosh(\sqrt{s})} - \frac{1}{s} \nonumber \\
    &=& -\frac{1}{2}.
\end{eqnarray}
Substituting into Eq.~(\ref{nondim_int_eq}) gives
\begin{equation}
    \overline{\Psi}_{\DT,0} = V_0 + \frac{1}{2},
\end{equation}
which follows that
\begin{eqnarray}
    \overline{\Psi}_\DT = \frac{V}{\kappa} + \frac{X_\hit^2}{2D}.
\end{eqnarray}
Therefore, the asymptotic approximation of the mean particle transport time along protrusion is
\begin{eqnarray} \label{PTT_largeV}
    \overline{\psi}_\DT(V) = \frac{V}{\kappa} + \frac{\mathbb{E}[X_\hit^2]}{2D}.
\end{eqnarray}
Moreover, the integral term in Eq.~(\ref{LNA_DT}) is non-positive, and so we have a lower bound
\begin{equation} \label{LBD_DT}
    \Psi_\DT \geq \frac{V}{\kappa}.
\end{equation}
Numerical comparison in Fig.~\ref{fig6}(b) confirms that this is an approximation for large $V$.

\respI{Finally, important model parameters and variables for the direct transport model are summarized in Table.~\ref{tab1}, which appear in the next sections.}

\begin{table}[t!] {\color{black}
\caption{\label{tab1} Important parameters and variables for direct transport mechanisms. Single protrusion event ({\it top}), multiple protrusion events ({\it middle}), and molecule transport along established protrusion ({\it bottom}).
}
\begin{ruledtabular}
\begin{tabular}{ccc}
Symbols\footnote{Tilde over a variable means the rare event approximation of the variable.}&
Meaning&
Mean \\
\colrule
$\rho_\DT$&Hitting probability&-\\
$X_\hit$&Protrusion length when hitting target&$\lambda_\hit$\\
$X$&Unconditional protrusion length&$\lambda$\\
$T_\hit$&Protrusion time when hitting target&$\tau_\hit$\\
\colrule
$\kappa_\DT$&Protrusion projection rate&-\\
$\mathcal{X}$&Total polymerization length&$\xi$\\
$\mathcal{T}_\DT$&First passage time to target&$\tau_\DT$\\
\colrule
$\kappa$&Molecule synthesis rate&-\\
$V$&Number of transporting molecules&-\\
$\Psi_\DT$&Molecule transport time along protrusion&$\psi_\DT$
\end{tabular}
\end{ruledtabular}}
\end{table}

\section{Diffusive particle transport under degradation} \label{Sect3}
We next consider the classical intercellular signaling mechanism of diffusion with degradation, which is referred to as ``mortal'' diffusion \cite{Lawley2021SIAP,Meerson2015PRL}.
The ``mortal'' molecules eventually either hit the target or are degraded; we denote the portion of hitting the target by $\rho_\diff$. Together with the hitting probability, we again would like to determine the transport time $\psi_\diff$ for delivering the required number of signaling molecules $V$, in terms of the model parameters. Furthermore, as we did in the previous section, we will approximate the transport time in the regime of large $V$.

We consider a source where signaling particles are released at a constant rate, which then diffuse in two dimensions to find target $\Omega \subset \mathbb{R}^2$. To make a ``fair'' comparison, we assume that the molecule production rate $\kappa$ and the diffusivity $D$ are the same as the diffusion along with the (1D) protrusion in the direct transport model. We also assume that molecules survive for an exponential amount of time with rate $g$. \respI{In this section, $u(x,t)$ again denotes the molecule concentration at position $x$ and time $t$, but in two-dimensional domain.} 
\respI{That is, the concentration variable has the unit of inverse square length.} 
This concentration satisfies
\begin{equation} \label{diff_eq}
    \frac{\partial u}{\partial t} = D\nabla^2 u - gu + \kappa \delta(x-x_s),
\end{equation}
where $x_s$ is the location of the source (and $\delta$ denotes the delta function in this section, not the difference of a function between the domain). This equation is supplemented by the absorbing boundary condition
\begin{equation}
    u(x \in \partial\Omega, t) = 0,
\end{equation}
where $\partial \Omega$ is the boundary of $\Omega$. If $\mathbb{R}^2 \backslash\Omega$ is unbounded, it requires an asymptotic boundary condition
\begin{equation}
    \lim_{\|x\| \to 0} u(x,t) = 0.
\end{equation}
Then the flux to the target\respI{, which is the unit of inverse time,} satisfies
\[
    J_\diff(t) = -D \int_{\partial \Omega} \frac{\partial u(x,t)}{\partial n} da, 
\]
where $\partial/\partial n$ represents the inward normal derivative to target $\Omega$. Since the flux converges to the steady-state, the hitting probability should satisfy
\[
    J_\diff^\ss  = \kappa \rho_\diff.
\]
Similar to the diffusion along a protrusion, we can also define the transport time for delivering a required number of molecules $V$ by
\begin{equation} \label{int_eq2}
    \int_0^{\psi_\diff} J_\diff(t') dt' = V.
\end{equation}

\begin{figure}[t!]
\includegraphics[width=8.6cm]{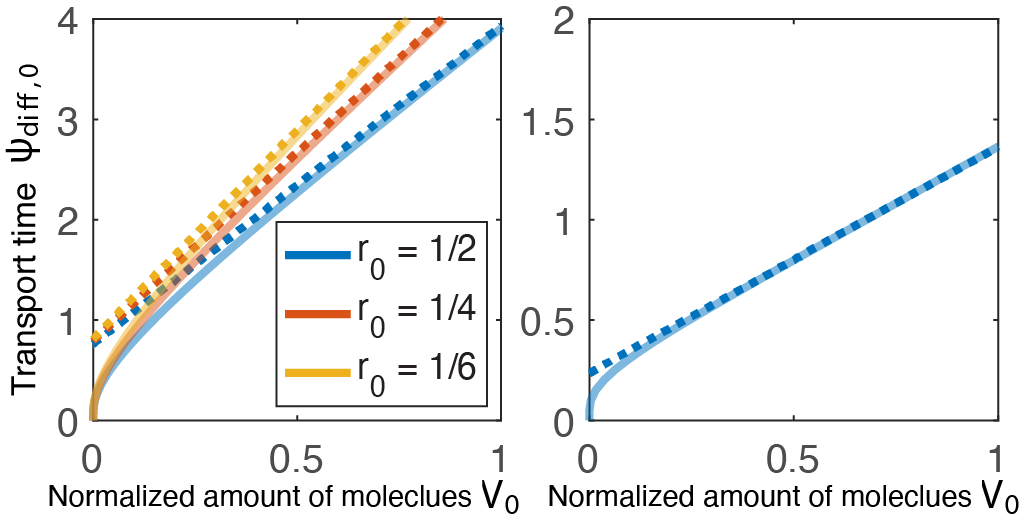}
\caption{\label{fig7} \respI{Non-dimensional} molecule transport time of the diffusion model $\psi_{\diff,0}$ to the target as a function of non-dimensional particle number $V_0$ in case of the disk shape (\textit{left panel}) and the annulus shape (\textit{right panel}) with various non-dimensional radius $r_0$. Exact transport time (\textit{solid curve}) is compared with the asymptotic approximation (\textit{dotted curve}). Non-dimensional degradation rate is chosen by $d_0 = 1/2$.}
\end{figure}

We determine the transport time by taking the Laplace transformation. We first non-dimensionalize using the variables $x_0 = x/d$, $t_0 = tD/d^2$, $V_0 = VD/(\kappa d^2)$, $J_{\diff, 0} = J_\diff/\kappa$. The only difference from the previous section is that the length is non-dimensionalized by the minimum distance $d$ between the source and the target. Then the implicit equation for $\psi_\diff$ becomes
\begin{equation}
    \mathcal{V}_\diff(\psi_{\diff,0}) := \int_0^{\psi_{\diff,0}} J_{\diff,0} (t_0') dt_0' = V_0,
\end{equation}
where $\mathcal{V}_\diff(t_0)$ is again the non-dimensional number of transported particles to the target via diffusion over non-dimensional time $t_0$. Another Laplace transformation yields
\begin{equation}
    \mathcal{L}[\mathcal{V}_\diff](s) = \frac{1}{s} \mathcal{L}[J_{\diff, 0}](s),
\end{equation}
which depends on the non-dimensional target radius $r_0 = r/d$ and degradation rate $g_0 = g d^2/D$. This allows finding the asymptotic flux
\begin{eqnarray*}
    J_{\diff,0}^\ss &=& \lim_{s \to \infty} s \mathcal{L}[J_{\diff,0}](s) \\
    &=& \rho_\diff,
\end{eqnarray*}
which is the same as the hitting probability of a single diffusive particle.
Another numerical inversion of the Laplace transformation gives $\mathcal{V}_\diff (t)$. Then the particle transport time takes the form  
\begin{equation}
    \psi_\diff(V) = \frac{d^2}{D}\mathcal{V}_\diff^{-1} \left( \frac{DV}{\kappa d^2}\right).
\end{equation}
We approximate the particle transport time for large $V$ by
\[
    \int_0^\infty J_{\diff,0}(t_0') - J_{\diff,0}^\ss dt_0' + J_{\diff,0}^\ss \overline{\psi}_{\diff,0} = V_0,
\]
which follows that
\begin{equation} \label{Psidiff}
    \overline{\psi}_{\diff,0}(V_0) = \frac{V_0}{\rho_\diff} + \frac{1}{2}\Psi_\diff,
\end{equation}
where
\[
    \Psi_\diff = 2\int_0^\infty 1 - \frac{J_{\diff,0}(t_0')}{J_{\diff,0}^\ss} dt_0'.
\]
Therefore, we have
\begin{equation}
    \overline{\psi}_\diff = \frac{V}{\kappa \rho_\diff} + \frac{d^2}{2D}\Psi_\diff.
\end{equation}
The exact formulation and detailed derivation can be found in \cite{SM}. 
\respI{Our analysis can be extended to the corresponding 3D model with the Laplace operator in the spherical coordinate.} Numerical simulation in Fig.~\ref{fig7} shows that the approximation agrees with the direct inversion for a large number of molecules.

\section{Utility Analysis} \label{Sect4}
How do we quantify the ``efficiency'' of an intercellular signaling transport mechanism? One important consideration is the energy cost for transporting signaling molecules from a source cell to a target cell. The cost arises from synthesizing signaling molecules and polymerizing filaments for cellular protrusions. If we consider only the energy aspect, it would give an absurd prediction for optimal parameters. For example, the energy cost for contact formation, which is proportional to the total polymerization length in (\ref{lambda_tau}), is minimized when the projection rate is zero -- even though, in reality, cells extend multiple protrusions to find the target \cite{Kornberg1999,Kornberg2017}. Another important consideration is the benefit accrued by cells if they communicate quickly. Thus, another crucial aspect of efficiency is the time required for transporting a required amount of signaling particles.

Under competition between energy cost and \respI{communication speed}, one natural measure of the signaling efficiency is their ratio
\[
    \gamma = \mathbb{E}\left[\frac{\mathcal{E}}{\gamma_0 \mathcal{T}^{-1}}\right],
\]
where $\gamma_0$ is a conversion factor from transport time to benefit. In this formulation, we assume the benefit accrued by more rapid signaling is linear in the rate of signaling transmission, although in some biological situations benefits of rapid signaling may be non-linear.  The resulting cost-benefit ratio is then
\begin{equation} \label{CBR}
    \gamma = \gamma_0^{-1} \mathbb{E}[\mathcal{E}\mathcal{T}].
\end{equation}
Aside from this cost-benefit ratio, we might alternatively consider a utility function given by the total energetic cost of a successful signal
\begin{equation} \label{TEF}
    \eta = \mathbb{E}[\mathcal{E}] + \eta_0\mathbb{E}[\mathcal{T}],
\end{equation}
where the second term penalizes a long signaling process due to the energetic cost of maintaining cell homeostasis during the signaling search process (the factor $\eta_0$ denotes a conversion between time and energetic cost of homeostasis). In this section, we determine and compare the cost-benefit ratio of the transport mechanisms. In particular, we show that the direct transport is optimized at some protrusion projection rate \respI{$\kappa_\DT^*$ (which subscript will be modified depending on the cost-benefit function)}. Then we extend the qualitative results for the cost-benefit ratio as a measure of efficiency based on total energetic cost, \respI{which allows determining the critical condition balancing the efficiency of the two mechanisms, especially the critical number of transporting molecules $V_{c}$.}

\subsection{Efficient contact formation at optimal projection rate} \label{sect4A}
\begin{figure}[t!]
\includegraphics[width=8.6cm]{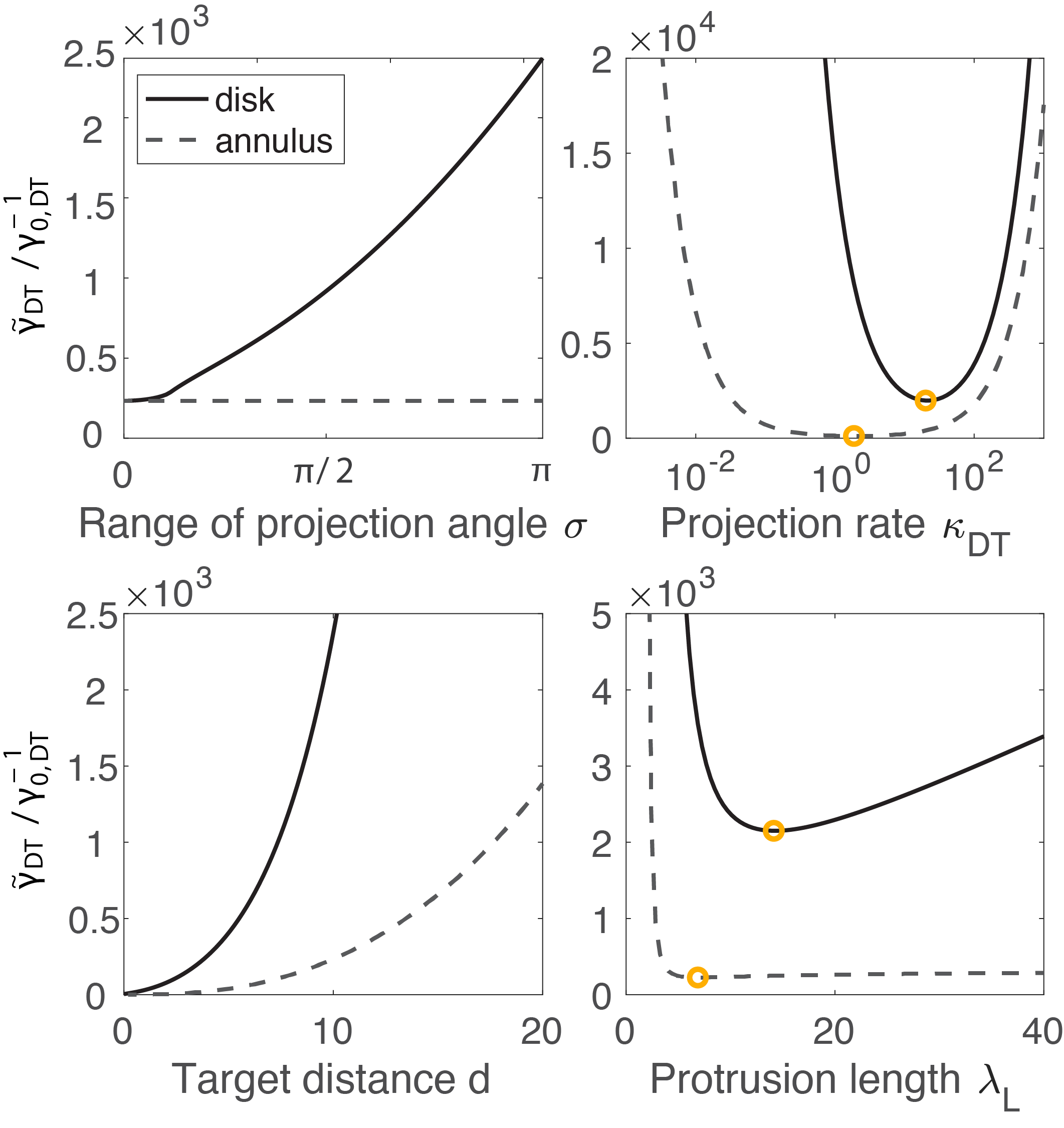}
\caption{\label{fig8} Cost-benefit ratio of contact formation by protrusions with various parameters in case of the disk-shaped target (\textit{solid curve}) and the annulus-shaped target (\textit{dotted curve}). There exist an optimal projection rate and mean protrusion length (\textit{yellow circle}). Parameters are chosen as follows: $d = 10$ $\mu$m and the others are the same as Fig.~\ref{fig5}.}
\end{figure}

First, we consider the contact formation process of the direct transport model, temporarily neglecting molecule transport following contact formation.
Let $\Delta_{\DT}$ denote by the energy cost of protrusion polymerization per unit length. Then the total energy cost for contact formation is proportional to the total polymerization length
\[
	\mathcal{E}_{\DT} = \Delta_{\DT} \mathcal{X},
\]
assuming that the energy is not required for depolymerization. Together with the contact formation time $\mathcal{T}_\DT$, the cost-benefit ratio of direct contact formation is then 
\[
	\gamma_{\DT} = \mathbb{E}[\mathcal{E}_{\DT} \mathcal{T}_{\DT}] = \gamma_{0,\DT}^{-1} \mathbb{E}[\mathcal{X} \mathcal{T}_{\DT}],
\]
where $\gamma_{0,\DT}^{-1} = \gamma_0^{-1} \Delta_{\DT}$. \respI{The rare-event approximation of $\gamma_\DT$ takes the form
\begin{equation} \label{gamma_DT}
	\frac{\widetilde{\gamma}_{\DT}}{\gamma_{0,\DT}^{-1}} = \frac{c_{-1}}{\rho_\DT^2\kappa_\DT} + c_1 \lambda \kappa_\DT  + c_0.
\end{equation}
The exact expression of coefficients and derivation can be found in \cite{SM}. 
This approximation can be interpreted as follows:  
\respII{When the projection rate is slow (and the first term dominates), the contact formation process can be more efficient with a faster projection rate by boosting the search process; When the projection rate is fast (and the second term dominates), it can be less efficient with a faster projection rate because redundant protrusions are made.}
Therefore, there exists an optimal projection rate $\kappa_\DT^\star$ that is neither too slow to find the target nor wastes too many protrusions. The cost-benefit ratio as a function of $\kappa_\DT$ is depicted in the top-right panel of Fig.~\ref{fig8}. Note that the corresponding panel is a semi-logarithmic plot and so the curve appears like an exponential curve.}

Fig.~\ref{fig8} shows the cost-benefit ratio of the contact formation, in response to variation in other parameters. $\widetilde{\gamma}_\DT$ is monotonic with respect to $\sigma$ and $d$ \respI{({\it left panels})}. That is, as the range of projection angle is tighter, the source cell can hit the target cell in a shorter time with a smaller number of protrusions. Likewise, the cost-benefit ratio decreases when the target is closer. \respI{Bottom-right panel} also confirms that there is an optimal mean protrusion length (slightly longer than the target distance) that minimizes the cost-benefit ratio. \respI{Similar optimality (FPT vs. protrusion length) has been also observed in \cite{Bressloff2020PRE,Bressloff2021SIAP}.}

The cost-benefit ratio is non-monotonic with respect to the projection rate. The optimal projection rate can be found by solving the critical condition for $\kappa_{\DT}$, $0 = \partial \widetilde{\gamma}_{\DT}/\partial \kappa_{\DT}$. And so the optimal protrusion rate satisfies
\begin{equation} \label{opt1}
    \frac{1}{\kappa_\DT^\star \rho_\DT} =  \frac{\tau_\hit}{\chi_0},
\end{equation}
where $\chi_0 = Q\mathbb{E}[X_\hit]/\sqrt{\mathbb{E}[X_\hit^2]}$ \respI{and $\lambda Q^2 = \lambda + (1-\rho_\DT)\lambda_\miss$}.
In order to interpret this condition, we find the critical condition for the moment closure of the ratio, $0 = \partial (\widetilde{\tau}_\DT \widetilde{\lambda}_\DT)/\partial \kappa_\DT$, which yields
\begin{equation} \label{opt_mc}
    \frac{1}{\kappa_\DT \rho_\DT} = \tau_\hit.
\end{equation}
Here $\kappa_{\DT}^{-1}\rho_{\DT}^{-1}$ represents the time to generate a protrusion that hits the target, and $\tau_\hit$ is the protrusion traveling time to the target. These time intervals are balanced at the optimal rate of protrusion. \respI{Eq.~ (\ref{opt_mc}) assumes no correlation between the first passage time and the total polymerization length. Thus, $\chi_0$ in Eq.~(\ref{opt1}) is a correction term due to the correlation.}

\begin{figure}[t!]
\includegraphics[width=8.6cm]{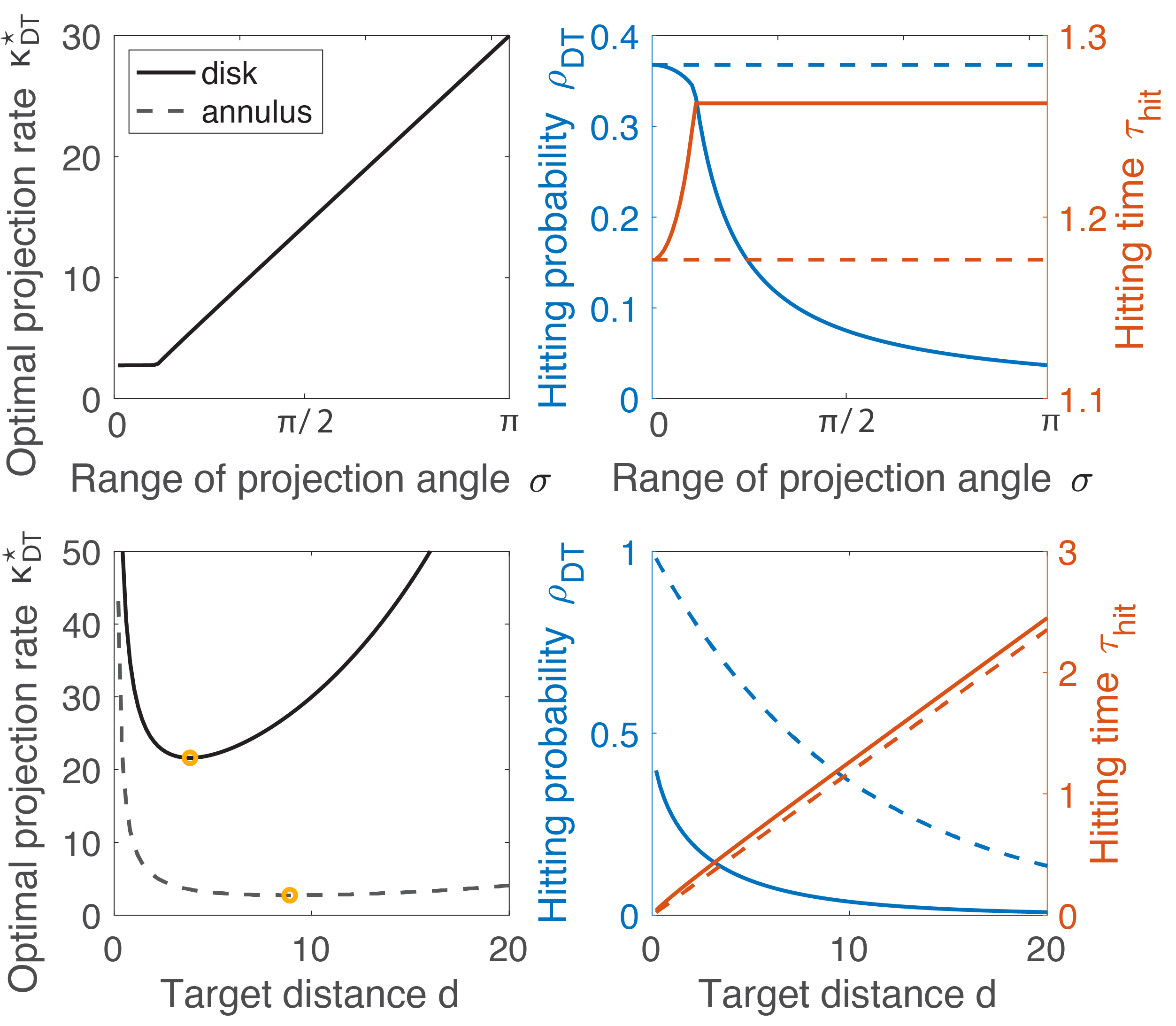}
\caption{\label{fig9} Optimal protrusion rate $\kappa_\DT^\star$ with various parameters, together with the statistics of the single protrusion event $\rho_\DT$ and $\tau_\hit$.  The optimal projection rate satisfies Eq.~(\ref{opt1}). Parameters are the same as Fig.~\ref{fig8}.}
\end{figure}
Numerical simulations in Fig.~\ref{fig9} show the optimal projection rate $\kappa_\DT^\star$, \respI{together with the hitting probability $\rho_\DT$ and the conditional hitting time $\tau_\hit$}, as a function of various parameters. \respI{We explain the behavior of $\kappa_\DT^*$ by $\rho_\DT$ and $\tau_\hit$, which determine $\kappa_\DT^\star$ by Eq. (\ref{opt1}).} 
The optimal projection rate monotonically decreases as the range of projection angle $\sigma$ is sharper. \respI{If the projection range is wider than the target, the conditional hitting time is constant. However, the hitting probability increases as the projection range is sharper. Thus, Eq. (\ref{opt1}) shows that $\kappa_\DT^*$ decreases as $\sigma$ is sharper.}
By contrast, the optimal projection rate does not change much over the target range $\sigma \in \phi(\Omega)$, because the \respI{unconditional hitting time ($\rho_\DT \tau_\hit$)} is nearly constant if the protrusions are always projected towards the target. However, the optimal projection rate has a non-monotonic behavior with respect to the target distance $d$. For a target that is farther from the source, the source cell prefers a faster projection rate because the target cell is harder to reach. The optimal projection rate is also fast when the target is close, even though the hitting probability is high, because the optimal rate balances the projection time for a hitting event and the travel time to the target cell (Eq. (\ref{opt1})). When the target is closer, the traveling time is shorter, which requires a shorter projection time for the hitting event by a faster projection rate.

\subsection{Optimal projection rate with particle transport}
Now we reconsider the cost-benefit ratio, including the process of particle transport along the protrusion following contact with a target. In addition to the contact formation time, it now takes more time to transport the required number of particles $V$ along the protrusion
\[
    \mathcal{T}_{\Sigma,\DT} = \mathcal{T}_\DT + \Psi_\DT(V;X_\hit).
\]
Since there is no loss of particles by the protrusion-mediated transport, the total energy cost is
\[
    \mathcal{E}_{\Sigma,\DT} = \mathcal{E}_\DT + \Delta_p V,
\]
where $\Delta_p$ is the energy price for synthesizing a single signaling particle. Therefore, the mean cost-benefit ratio becomes
\[
    \gamma_{\Sigma, \DT} = \gamma_0^{-1} \mathbb{E}[\mathcal{E}_{\Sigma,\DT} \mathcal{T}_{\Sigma,\DT}].
\]
\respI{We calculate the rare-event approximation (see \cite{SM} for details) 
and subtract the approximation with and without particle transport
\begin{equation} \label{DT_cbr} 
	\frac{\widetilde{\gamma}_{\Sigma, \DT}}{\gamma_{0,\DT}^{-1}} - \frac{\widetilde{\gamma}_{\DT}}{\gamma_{0,\DT}^{-1}} = c_{\Sigma,-1} \Delta_p V \kappa_\DT^{-1} + c_{\Sigma,1} \psi_\DT^{(1)}(V) \kappa_\DT + c_{\Sigma,0}.
\end{equation}
where $\psi_\DT^{(k)}(V) = \mathbb{E}[X_\hit^k \Psi_\DT(V;X_\hit)]$ for $k = 0,1,\cdots$.}
In contrast to Eq.~(\ref{gamma_DT}), there are now additional terms on the coefficients, which arise from the particle transport, including the energy cost for molecule production ($\Delta_p V$) and the particle transport time along the protrusion ($\psi_\DT^{(1)}(V)$).

Similar with the previous section, one can find the optimal projection rate \respI{$\kappa_{\Sigma, \DT}^*$} by solving the critical condition $0 = \partial \widetilde{\gamma}_{\Sigma, \DT} /\partial \kappa_\DT$. \respI{That is, $\kappa_{\Sigma,\DT}^*$ optimizes the cost-benefit ratio of the whole direct transport process, including contact formation by protrusions and particle transport along the protrusion linkage. In contrast, $\kappa_\DT^*$ only optimizes the cost-benefit ratio of the contact formation process. We compare those optimal projection rates by calculating their ratio
\begin{equation}
	\left(\frac{\kappa_{\Sigma, \DT}^\star}{\kappa_\DT^\star}\right)^2 = \frac{c_{-1} \rho_\DT^{-2} + c_{\Sigma,-1}\Delta_p V}{c_1 \lambda + c_{\Sigma,1} \psi_\DT^{(1)}(V)}.
\end{equation}}
This yields the following equivalent condition
\begin{equation} \label{crit_ER}
	\kappa_{\Sigma, \DT}^\star > \kappa_{\DT}^\star  ~\Longleftrightarrow~ \frac{\varepsilon_p}{\varepsilon_\DT} > \chi_\varepsilon(V),
\end{equation}
where the energy rates satisfy 
\[
    \varepsilon_p = \kappa \Delta_p, \qquad \varepsilon_\DT = \Delta_\DT \lambda \kappa_\DT^\star,
\]
and the critical energy ratio is
\[
    \chi_\varepsilon(V) = \frac{\chi_0 \kappa \psi_\DT^{(1)}(V)}{ \lambda_\hit V}.
\]
Here $\varepsilon_p$ represents the energy rate for signaling molecule production and Eq.~(\ref{lambda_tau}) shows that $\varepsilon_\DT$ is the average energy rate for the contact formation without particle transport. Therefore, one interpretation of inequality Eq.~(\ref{crit_ER}) is that, for given $V$, if the particle synthesis energy rate is sufficiently larger than the average contact formation energy rate, cells prefer the faster projection rate to be optimized. Otherwise, the slower protrusion rate is more efficient.

We are also interested in the cost-benefit ratio, and optimal projection rate, in the limit of a large number of transporting particles, $V$. Using the fact that
\begin{equation} \label{trt_bdd}
    \frac{V}{\kappa} \leq \Psi_\DT(V;X_\hit) \leq\frac{V}{\kappa} + \frac{X_\hit^2}{2D},
\end{equation}
we have the limit
\begin{equation}
    \lim_{V \to \infty} \chi_\varepsilon (V) = \chi_0.
\end{equation}
Therefore, the asymptotic critical condition is
\begin{equation} \label{acrit_ER}
	\kappa_{\Sigma, \DT}^\star > \kappa_{\DT}^\star  ~\Longleftrightarrow~ \frac{\varepsilon_p}{\varepsilon_\DT} > \chi_0,
\end{equation}
in the limit of a large number of transporting particles. In particular, we can extend the asymptotic condition to any number of particles
\begin{equation} \label{DTcrit1}
	\frac{\varepsilon_p}{\varepsilon_\DT} < \chi_0  ~\Longrightarrow~ \kappa_{\Sigma, \DT}^\star < \kappa_{\DT}^\star
\end{equation}
by the following inequality derived from Eq.~(\ref{LBD_DT})
\[
    \chi_\varepsilon(V) \geq \chi_0,
\]
for any $V$. In other words, the optimal projection rate with particle transport is always slower than the optimal rate without particle transport, if the molecule synthesis cost is sufficiently cheap compared to protrusion elongation cost, regardless of $V$.

\begin{figure}[t!]
\includegraphics[width=8.6cm]{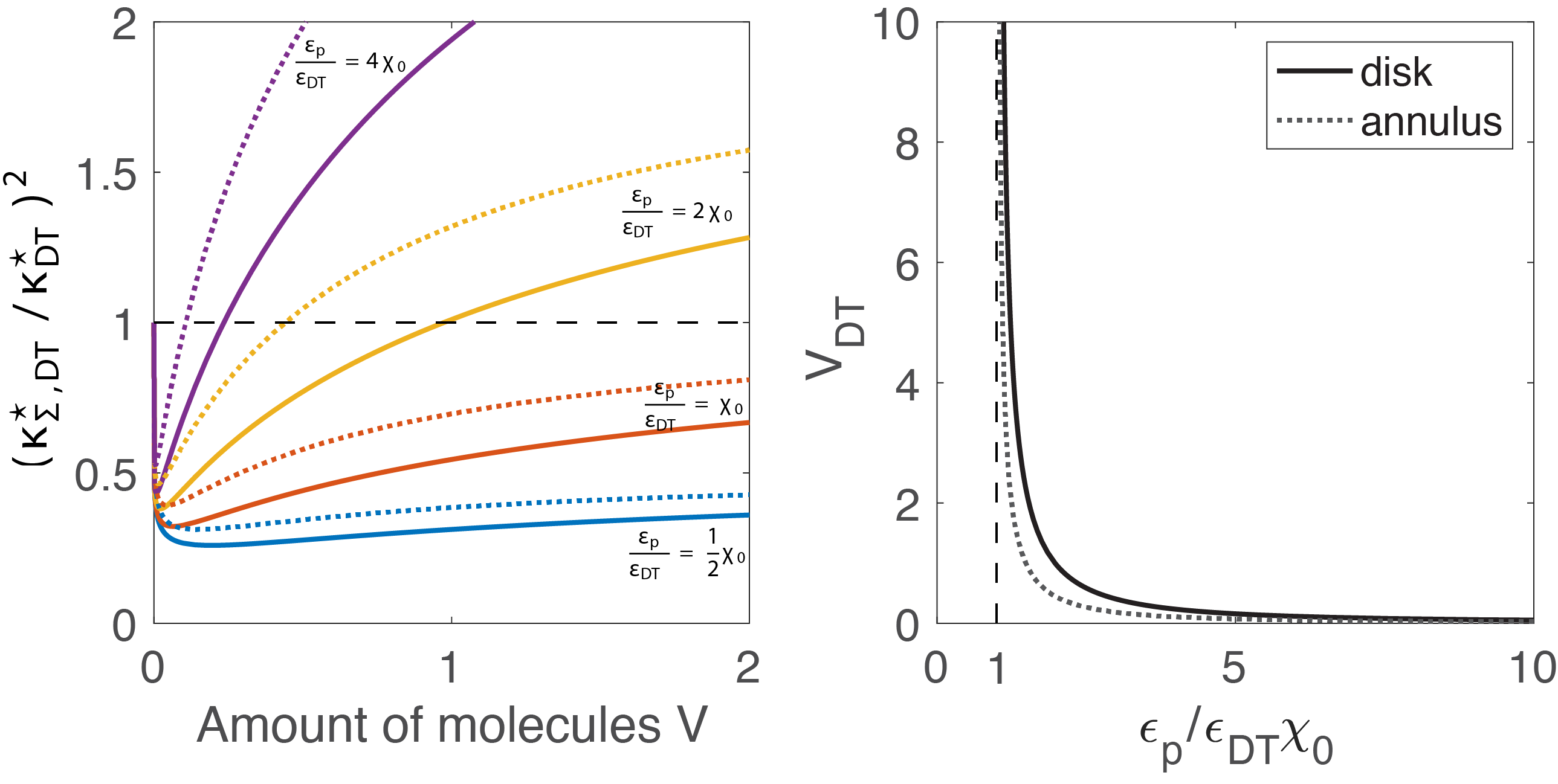}
\caption{\label{fig10} Comparison between the optimal projection rate with and without the particle transport. Squared fraction of the optimal projection rates $(\kappa_{\Sigma,\DT}^\star/\kappa_\DT^\star)^2$ as a function of required number of transporting particles $V$ with various fraction of energy rate (\textit{left panel}) in Eq.~(\ref{DTcrit1}). Critical number $V_{\DT}$ in Eq.~(\ref{DTcrit2}) as a function of the fraction of the energy rates $\varepsilon_p/\varepsilon_\DT \chi_0$ (\textit{right panel}). Parameters $\kappa, \Delta_p, \Delta_\DT$ are chosen according to the value of $\varepsilon_p/\varepsilon_\DT \chi_0$. Other parameters are the same as Fig.~\ref{fig8}.}
\end{figure}

However, this strict ordering of optimal \respII{projection rates} with or without particle transport in Eq.~(\ref{DTcrit1}) does not hold, but depends on $V$ when the energy for producing signaling molecules is sufficiently large compared to the energy for contact formation, as shown in Fig.~\ref{fig10}. 
In other words, there exists a number $V_{\DT}$ such that 
\begin{equation} \label{DTcrit2}
    \frac{\varepsilon_p}{\varepsilon_\DT} > \chi_0 ~\Longrightarrow~
	\begin{cases}
		\kappa_{\Sigma,\DT}^\star < \kappa_\DT^\star, & V<V_{\DT}\\
		\kappa_{\Sigma,\DT}^\star > \kappa_\DT^\star, & V>V_{\DT}\\
	\end{cases}.
\end{equation}
This statement can be shown by using a standard regular perturbation argument. 
Expanding $\Psi_\DT(V) = \Psi_{\DT,0} V^\beta + \cdots$ and substituting into Eq.~(\ref{int_eq}), we determine the leading order $\beta = 1/2$. This follows that
\begin{equation}
    \chi_\varepsilon(V) \sim V^{-1/2},
\end{equation}
for small $V$.
Therefore, for any energy cost ratio $\varepsilon_p/\varepsilon_\DT$, there exists small $V_{\DT}>0$ such that $\chi_\varepsilon(V) > \varepsilon_p/\varepsilon_\DT$ for all $V < V_{\DT}$. 

One interesting observation is that the ratio of optimal projection rates with or without particle transport is non-monotonic in $V$, as shown in Fig.~\ref{fig10}. This arises from the effective transport behavior of diffusion along with a 1D domain. At large $V$, the fluxes at both ends of the protrusion converge to the maximum, and the transport behaves like an advective process. Thus, to minimize the cost-benefit ratio, the optimal rate of protrusion is faster when accounting for the transport process, because the protrusion elongation cost is relatively cheap. However, at small $V$, particles are transported more like diffusion, with a longer traveling time. This particle transport time dominates the cost-benefit ratio so that the protrusion elongation cost becomes relatively large, and the optimal rate of protrusion is slower when accounting for particle transport.

\subsection{Signaling by direct transport versus diffusion}
We can form expectations for when a cell would evolve to use direct transport or, alternatively, mortal diffusion for signaling, by inspecting the cost-benefit ratio in each case. To do this we now analyze the cost-benefit ratio under mortal diffusion. Due to the degradation of signaling molecules, the source cell is required to release $V/\rho_\diff$ particles to deliver $V$ to the target cell. Therefore, the total energy cost under mortal diffusion is
\[
    \mathcal{E}_{\Sigma,\diff} = \frac{\Delta_p V}{\rho_\diff},
\]
which takes the following amount of time
\[
    \mathcal{T}_{\Sigma,\diff} = \psi_\diff(V).
\]
And so the cost-benefit ratio under the diffusion mechanism takes the form of
\begin{eqnarray}
    \frac{\gamma_{\Sigma,\diff}}{\gamma_0^{-1}} &=& \mathcal{E}_{\Sigma,\diff}\mathcal{T}_{\Sigma,\diff} \nonumber \\
    &=& \frac{\Delta_p V \psi_\diff(V)}{\rho_\diff}. \label{diff_cbr}
\end{eqnarray}

Finally, we can compare the cost-benefit ratio of two signaling mechanisms. \respI{One important result is that the direct transport outperforms diffusion for a sufficiently large number of molecules V, as shown in Fig. \ref{fig11}(a).
\begin{figure}[t!]
\includegraphics[width=8.6cm]{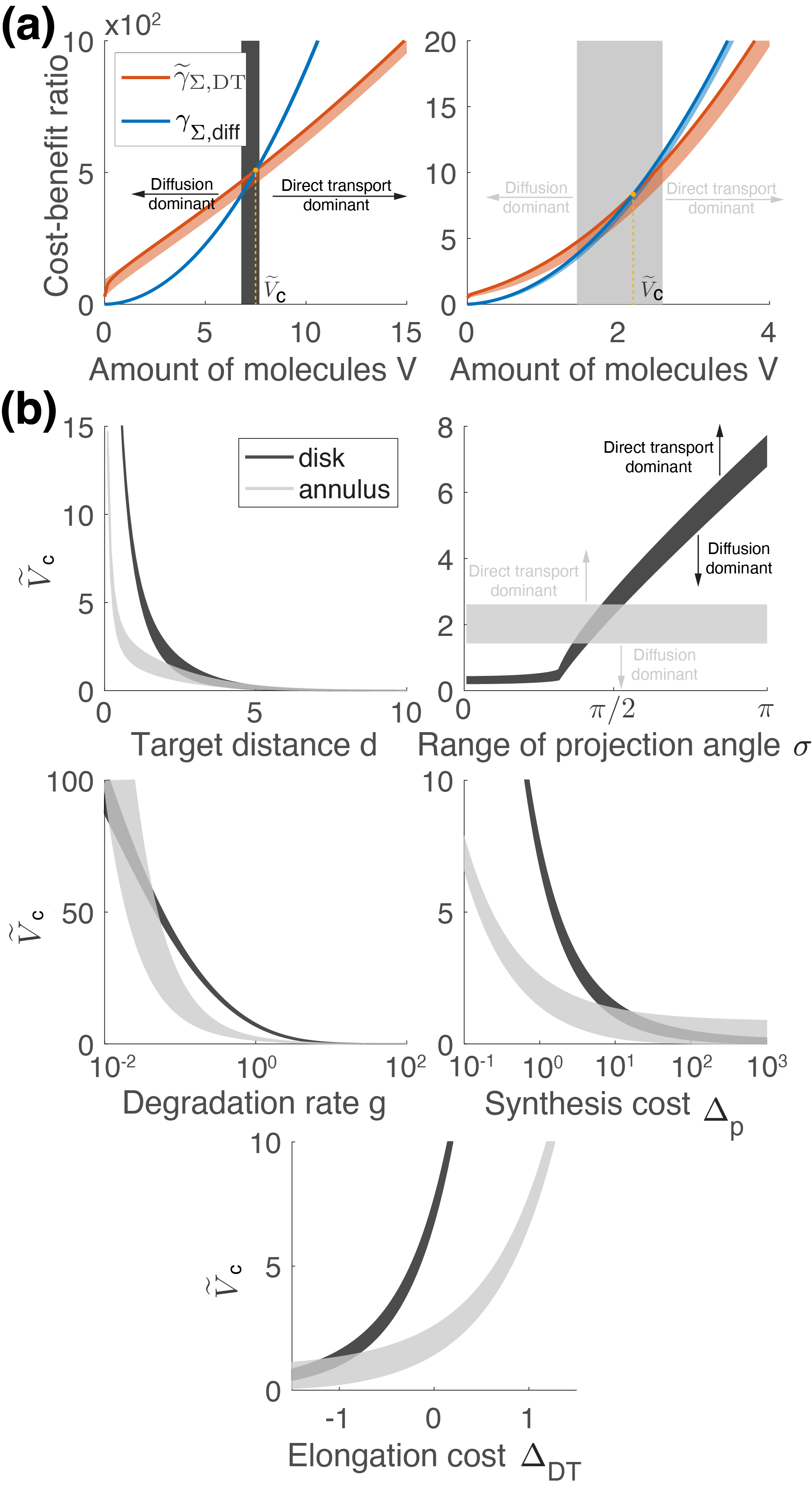}
\caption{\label{fig11} Critical amount of transporting molecules. (a) Critical number $\widetilde{V}_{\Sigma,c}$ (\textit{yellow dot}) intersecting the cost-benefit ratio of the direct transport model $\widetilde{\gamma}_{\Sigma,\DT}$ (\textit{red curve}) and the diffusion model $\gamma_{\Sigma,\diff}$  (\textit{blue curve}) for the disk-shaped target (\textit{left panel}) and the the annulus-shaped target (\textit{right panel}). Upper and lower bound for $\widetilde{V}_{\Sigma,c}$ (\textit{gray area}) can be determined by the bounds for $\widetilde{\gamma}_{\Sigma,\DT}$ (\textit{red area}) and $\gamma_{\Sigma,\diff}$ (\textit{blue area}). If the required number of transporting particles is larger than the critical number, then the direct transport model is more efficient. Otherwise, the diffusion model is preferred. (b) Plots of the critical number with various parameters. Parameters are as follows: $d = 1$ $\mu\text{m}$, $D = 1$ $\mu\text{m}/\text{min}^2$, $\kappa = 1/\text{min}$, $g = 1/\text{min}$, $\Delta_\DT = 1 \varepsilon/\mu\text{m}$, $\Delta_p = 1 \varepsilon$, where $\varepsilon$ is the required number of ATP for synthesizing a single signaling molecule. Others are the same as Fig.~\ref{fig8}.}
\end{figure}
This observation can be mathematically shown by the asymptotic behavior of the cost-benefit ratio for large $V$.}
At large $V$, Eqs. (\ref{DT_cbr}) and (\ref{diff_cbr}) indicate that the leading order behavior of the ratios are 
\begin{equation} \label{asymVlarge}
    \frac{\gamma_{\Sigma,\DT}}{\gamma_0^{-1}} \sim \frac{\Delta_p V^2}{\kappa}, \qquad \frac{\gamma_{\Sigma,\diff}}{\gamma_0^{-1}} \sim \frac{\Delta_p V^2}{\kappa \rho_\diff^2}.
\end{equation}
Since $\rho_\diff < 1$, then $\gamma_{\Sigma,\DT} < \gamma_{\Sigma,\diff}$ for sufficiently large $V$. That is, if the cell must deliver a large number of signaling molecules, the direct transport mechanism is cheaper than the diffusion mechanism, because the contact formation energy cost per transported particle is cheaper.

\respI{Furthermore, we show that there exists a critical number of molecules $V_c$ by showing the other end of the asymptotic behavior of the cost-benefit ratio.} At small $V$, the cost-benefit ratios converge to
\begin{equation} \label{comp_zero}
    \lim_{V \to 0} \gamma_{\Sigma,\DT}(V) = \gamma_\DT, \qquad \lim_{V \to 0} \gamma_{\Sigma,\diff}(V) = 0.
\end{equation}
This follows that $\gamma_{\Sigma,\DT} > \gamma_{\Sigma,\diff}$ for sufficiently small $V$. In this limit, due to the cost of contact formation, the diffusion mechanism is always cheaper than direct transport. Together with Eq. (\ref{asymVlarge}), this guarantees the existence of the critical number of particles $V_c$ that determines whether diffusion or direct transport is preferable, namely:  
\begin{equation} \label{cr_cond}
    \gamma_{\Sigma,\DT}(V_c) = \gamma_{\Sigma,\diff}(V_c).
\end{equation}
Numerical simulation in Fig.~\ref{fig11}(a) confirms the existence of the critical number of particles above which direct transport is preferred and below which diffusion is preferred. 

\respI{To study the behavior of the critical number in response to various parameters, we use upper and lower bounds for the rare-event approximation of $V_c$. We first define the rare-event approximation of $V_c$ by solving
\begin{equation}
	\gamma_\Sigma(\widetilde{V}_c) := \gamma_{\Sigma,\diff}(\widetilde{V}_c) - \widetilde{\gamma}_{\Sigma,\DT}(\widetilde{V}_c) =0.
\end{equation}
We then determine bounds for $\widetilde{V}_c$ by using the analytic bounds for the transport time in Sect. \ref{sect2D} and \ref{Sect3}.} 
Denote $\overline{\psi}$ ($\underline{\psi}$) by the upper (lower) bound for the transport time $\psi$ and denote $\gamma(V,\psi)$ by the cost-benefit ratio as a function of $\psi$. Then the cost-benefit ratio is bounded by
\begin{equation}
    \gamma(V,\underline{\psi}) \leq \gamma(V) \leq \gamma(V,\overline{\psi}).
\end{equation}
\respI{Thus, the difference $\gamma_\Sigma$ is bounded by} 
\begin{equation}
    \underline{\gamma}_\Sigma(V) \leq \gamma_\Sigma(V) \leq \overline{\gamma}_\Sigma(V)
\end{equation}
where $\underline{\gamma}_\Sigma(V) = \gamma_{\Sigma,\diff}(V,\underline{\psi}_\diff) - \gamma_{\Sigma,\DT}(V,\overline{\psi}_\DT^{(0)})$ and $\overline{\gamma}_\Sigma(V) = \gamma_{\Sigma,\diff}(V,\overline{\psi}_\diff) - \gamma_{\Sigma,\DT}(V,\underline{\psi}_\DT^{(0)})$. Therefore, the critical number of transport particles is bounded by
\begin{equation}
    \underline{V}_\Sigma \leq \widetilde{V}_{\Sigma,c} \leq \overline{V}_\Sigma
\end{equation}
where the bounds satisfy $\underline{\gamma}_\Sigma(\overline{V}_\Sigma) = 0$ and $\overline{\gamma}_\Sigma(\underline{V}_\Sigma) = 0$.

Numerical simulation in Fig.~\ref{fig11}(b) shows the behavior of the critical transport size as a function of various parameters by plotting the upper and lower bounds. 
If the target is far from the source cell, then the direct transport mechanism is generally preferred, unless the required number of transport molecules is very small (that is, $\widetilde{V}_{\Sigma,c}$ is small), because the protrusion search process is cheaper than diffusive particle synthesis. If the projections are aimed at the target more precisely, then again the direct contact model is typically preferred ($\widetilde{V}_{\Sigma,c}$ small) because the contact formation cost is small. In the case of an annular target, the critical number is independent of $\sigma$ because the target distance is uniform for any projection angle. At small degradation rates, the hitting probability of diffusive particles is higher and so diffusion is typically preferred ($\widetilde{V}_{\Sigma,c}$ large). 
When the cost of particle synthesis is cheap enough, diffusion is generally preferred ($\widetilde{V}_{\Sigma,c}$ large), especially so for a disk-shaped target because it has a smaller hitting probability. And for large elongation costs, the diffusion model is typically preferred ($\widetilde{V}_{\Sigma,c}$  large, especially for a disk-shaped target).

One natural question is whether, for any set of parameters, direct transport can always dominate diffusion, regardless of the size of required transport molecules, $V$. More precisely, we would like to find a bounded set of parameters such that
\[
     \gamma_\Sigma(V) \geq 0,
\]
for all $V$. Here we introduce the rare-event approximation for analytic simplicity. We choose $\alpha_1 = \Delta_\DT$ and $\alpha_2 = v_+^{-1}$ as the parameters for the dominant condition, then the cost-benefit ratio takes the form
\begin{equation}
    \gamma_\Sigma(\alpha_1,\alpha_2,V) = \varphi_0(V) - \alpha_1 \varphi_1(V) - \varphi_2(\alpha_2;V), 
\end{equation}
where 
\[
    \varphi_0(V) = \Delta_p V \left( \frac{\psi_\diff(V)}{\rho_\diff} - \psi_\DT^{(0)}(V)\right),
\]
$\varphi_1(V) = \mathcal{R}_{\Sigma,\DT}(V) - \mathcal{R}_\DT$, and $\varphi_2(\alpha;V)/2 = \alpha \text{Var}[X_\hit^2]\lambda_\hit \lambda_0/\rho_\DT + \rho_\DT^{-2}(\lambda_0 Q^2 + \Delta_p \rho_\DT V/\Delta_\DT) \times (\alpha^2 \mathbb{E}[X_\hit^2] + \alpha \psi_\DT^{(1)}(V))^{1/2}$.
We want to find non-zero $\alpha_1, \alpha_2$ such that there is a lower bound $\gamma_\Sigma(V) \geq \gamma_{\Sigma,0} > 0$ for all $V$. If $\varphi_0(V) \leq 0$ for some interval including $V = 0$, then $\gamma_\Sigma(V) < 0$ over the interval. Thus, one cannot find any non-zero parameters. In other words, the diffusion mechanism dominates the direct transport mechanism over the interval. Although $\varphi_0(V) \geq 0$ for all $V$, we still cannot find the set of parameters because 
\begin{equation}
    \lim_{V \to 0} \gamma_\Sigma (\alpha_1,\alpha_2,V) = \varphi_2(\alpha_2,0) < 0,
\end{equation}
for non-zero $\alpha_2$. This result implies that, for finite protrusion growth speed and the elongation cost, direct transport cannot dominate the diffusion mechanism for all $V$.

\subsection{Total energetic cost}
As an alternative to the cost-benefit ratio measure of signaling efficiency, we can also study the total energetic cost (\ref{TEF}) of direct transport and mortal diffusion. In general, the total energetic cost has qualitatively similar behavior to the cost-benefit ratio, such as an optimal projection rate and the critical number discussed in the previous section. We show that most of the qualitative properties of the cost-benefit ratio also hold for total energetic cost.

There exists an optimal projection rate of the total energetic cost. The total energetic cost of the contact form can be written by
\begin{equation}
    \eta_\DT = \Delta_\DT\mathbb{E}[\mathcal{X}] + \eta_0 \mathbb{E}[\mathcal{T}_\DT],
\end{equation}
whose rare-event approximation is
\begin{equation}
    \widetilde{\eta}_\DT = \Delta_\DT \widetilde{\xi} + \eta_0 \widetilde{\tau}_\DT.
\end{equation}
Substituting (\ref{t_DT_approx}) and (\ref{lambda_tau}) into the equation yields
\begin{equation} \label{etaDT}
    \widetilde{\eta}_\DT = \Delta_\DT \lambda \tau_\hit \kappa_\DT + \frac{\eta_0}{\kappa_\DT\rho_\DT} + \widetilde{\eta}_{\DT,0},
\end{equation}
where $\widetilde{\eta}_{\DT,0} = \Delta_\DT \lambda / \rho_\DT + \eta_0 \tau_\hit$. This is minimized at
\begin{equation}
    (\kappa_\DT^{-1}\rho_\DT^{-1})^2 = \frac{\Delta_\DT \lambda}{\eta_0 \rho_\DT} \tau_\hit,
\end{equation}
where the fraction on the right-hand side represents the energetic cost of maintaining metabolism during contact formation. That is, this optimal condition balances the time to generate a protrusion that will hit the target ($\kappa_\DT^{-1}\rho_\DT^{-1}$) and the geometric mean of the traveling time (converted into the energetic cost) for the protrusion to hit the target ($\tau_\hit$), which is analogous to Eq.~(\ref{opt1}). 
\respI{Numerical simulation (shown in Fig. S1 in \cite{SM}) shows that the total energetic function also has an optimal projection rate and the same qualitative behavior shown in Fig.~\ref{fig8}.}

We next consider the total energetic cost of the direct transport model with particle transport. Setting $\mathcal{E} = \mathcal{E}_{\Sigma,\DT}$ and $\mathcal{T} = \mathcal{T}_{\Sigma,\DT}$ gives
\begin{equation} \label{eta2}
    \eta_{\Sigma,\DT} = \eta_\DT + \Delta_p V + \eta_0 \psi_\DT^{(0)}(V).
\end{equation}
Since the second and the third terms are independent of the projection rate $\kappa_\DT$, the optimal condition with particle transport is the same as without particle transport
\begin{equation}
    0 = \frac{\partial \eta_{\Sigma,\DT}}{\partial \kappa_\DT} = \frac{\partial \eta_{\DT}}{\partial \kappa_\DT}.
\end{equation}
This implies that the optimal projection rate of the total energetic cost is independent of the number of transporting particles, which is qualitatively different behavior from the cost-benefit ratio in Eq.~(\ref{DTcrit2}).

The total energetic cost shows the same asymptotic behavior as the cost-benefit ratio with respect to the number of transporting particles. Similar to Eq.~(\ref{eta2}), one can define the total energetic cost of the diffusion mechanism
\begin{equation}
    \eta_{\Sigma,\diff} = \frac{\Delta_p V}{\rho_\diff} + \eta_0\psi_\diff(V).
\end{equation}
As $V \to 0$, diffusion is more efficient than direct contact
\begin{equation} \label{asymTEF1}
    \lim_{V \to 0} \eta_{\Sigma,\DT} = \eta_\DT, \qquad \lim_{V \to 0} \eta_{\Sigma,\diff} = 0, 
\end{equation}
and the opposite holds as $V \to \infty$ because
\begin{equation} \label{asymTEF2}
    \frac{\eta_{\Sigma,\DT}}{V} \sim \Delta_p  + \frac{\eta_0 }{\kappa}, \qquad \frac{\eta_{\Sigma,\diff}}{V} \sim \frac{1}{\rho_\diff}\left(\Delta_p + \frac{\eta_0 }{\kappa}\right)
\end{equation}
and $\rho_\diff < 1$. Therefore, there is a critical number satisfying $\eta_{\Sigma,\DT}(V) = \eta_{\Sigma,\diff}(V)$, which is analogous to Eq.~(\ref{asymTEF1}) and (\ref{asymTEF2}), that determines whether diffusion or direct transport is more efficient. 
\respI{The critical number for the total energetic cost has qualitatively similar behavior to that of the cost-benefit ratio analysis in Fig.~\ref{fig11} (see Fig. S2 in \cite{SM} for the numerical simulation).}

\section{Discussion} \label{Sect5}

In this paper, we have compared two qualitatively different mechanisms of intercellular signaling: direct transport (DT) and mortal diffusion. We first developed a protrusion-based model of direct transport, in which a source cell projects a series of protrusions in two dimensions until making contact with the target cell. Once contact is established, signaling molecules are then transported via diffusion along the one-dimensional protrusion. 
We calculated the mean effective protrusion length of a single protrusion, conditioned on either hitting or missing the target; and then used this to develop a rare-event approximation for the mean first passage time and mean total effective protrusion length in the case of multiple protrusions. Finally, we calculated the transport time for a required number of diffusive particles along an established protrusion. By contrast, in the case of diffusive signaling, we calculated the hitting probability and the transport time of diffusive molecules with degradation in two dimensions. We then introduced the cost-benefit ratio as a measure of the efficiency of these two mechanisms, comparing their relative efficiency across a range of parameters. We also compared the mechanisms for a different measure of efficiency, the total energetic cost, which shows qualitatively similar behavior to the cost-benefit ratio.

Two specific conditions that optimize and favor direct transport emerge from our analysis.
First, cellular protrusions that are accurately directed toward the target cell ($\sigma \ll \pi$) and whose length is similar to the target distance ($l \approx d$) minimize the cost-benefit ratio (as shown in Fig.~\ref{fig8} and Fig.~S1 in \cite{SM}). 
Indeed, theoretical analysis suggests that cells can gather accurate information about the location of a target cell by optimizing the distribution of receptors on the cell surface \cite{Miles2020PRL}. Therefore, in the case of bidirectional communications, we hypothesize that the DT mechanism may be efficiently utilized for responder cells. Empirical examples of this phenomenon are known to occur, as cells generate additional cytonemes by feedback signals following contact formation during FGF morphogen gradient formation in \textit{Drosophila} \cite{Du2018}.  
Second, the projection rate that optimizes contact formation, neglecting particle transport, balances the time required to initiate a successful protrusion with the time required for that protrusion to hit the target cell (Eq. \ref{opt1}). 

Interestingly, the optimal projection rates with and without particle transport have a nontrivial relationship. For a large number of transport particles, \respII{the condition} that the optimal projection rate with particle transport is faster than one without transport may have a trivial dependence on the energy cost of contact formation versus particle synthesis ($\varepsilon_p/\varepsilon_\DT$), as seen in Eq. (\ref{DTcrit1}). However, in the case of a small number of transport particles, the criteria have a non-linear relationship with the number of particles, as seen in Fig. \ref{fig10} and Eq. (\ref{DTcrit2}). This non-trivial dependence can be achieved through our mathematical analysis and simulation.

Our comparison of direct transport versus mortal diffusion highlights a critical number of required signaling molecules, $V_{c}$, that determines which of the two mechanisms is more efficient, for any given set of model parameters. In other words, which mechanism is more efficient will depend on the total number of required transport particles $V$, even when all other conditions are the same.  In particular, the direct transport model tends to be preferred over diffusion when the target cell is far from the source cell, across a very broad range of required transport molecules (as seen in Fig.~\ref{fig11} and Fig.~S2 in \cite{SM}). 
This theoretical result may help to explain why cytonemes and other signaling protrusions are often observed in long-range intercellular signaling \cite{Sanders2013,Gonzalez2019}.

\respII{By contrast, when the source and target cells are in close proximity -- eg, between niche cells and germ cells in \textit{Drosophila} testis \cite{Inaba2015} -- then direct transport is more efficient only when the required number of transport molecules is very large, under our model. And so this result provides a concrete prediction that can be tested experimentally for short-range signalling, if the amount of signaling molecules can be experimentally manipulated. That being said, the primary value of our model is not to produce precise quantitative predictions for experimental validation, but rather to determine which physical parameters are most important for governing whether direct transport versus diffusion will be the preferred mode of cell-cell signalling.}


A comparison between signaling by direct transport versus diffusion requires that we choose a measure of efficiency. Most of our qualitative conclusions hold for both of the two efficiency measures we have studied: a benefit-to-cost ratio of signaling speed to metabolic energy cost, as well as a metric based purely on energetic costs. Both utility functions predict an optimal projection rate and a critical number of required signaling molecules. 
Moreover, the critical number of signaling molecules has a similar dependence on variation in other model parameters, such as the distance to the target, information about the target location, the degradation rate of signaling molecules, and synthesis and elongation costs.

Our analysis has several limitations. We have assumed that signaling molecules degrade under the mortal two-dimensional diffusion model, whereas we assume they are protected against degradation when diffusing along the interior of a cellular protrusion. This assumption, which may be realistic in most settings, nonetheless penalizes the diffusion mechanism compared to direct transport. This affects the asymptotic behavior of the cost-benefit ratio at large $V$, that is, Eq.~(\ref{asymVlarge}) becomes
\begin{equation}
    \gamma_{\Sigma,\DT} \sim \frac{\Delta_p V^2}{\kappa \rho_{\diff,1}^2}, \qquad \gamma_{\Sigma,\diff} \sim \frac{\Delta_p V^2}{\kappa \rho_{\diff,2}^2},
\end{equation}
where $\rho_{\diff,k}$ is the hitting probability of signaling molecules in $k$-dimensional space to the target before degradation. For a disk target, 
$\rho_{\diff,1} > \rho_{\diff,2}$ \respI{(see \cite{SM} for proof)} 
and thus our qualitative results do not change if we allow the degradation of particles within the protrusion. However, we have the opposite relationship in the case of an annular target, which implies that the direct transport model can be monotonically less efficient than diffusion while incurring an additional cost for contact formation.
We have also assumed that signaling molecules and protrusions are nucleated from a point source. Indeed, the nucleation process can happen on the source cell surface, which can be approximated by introducing multiple point sources on the surface. This geometric effect can quantitatively affect our analytic results, but are expected to give qualitatively similar result when the overall rate of multiple sources is the same as the rate of the single source. 


There are many open questions and avenues for future research based on the simple modeling framework we have developed. One important area concerns the cost-benefit ratio as a measure of signaling efficiency. Our analysis has assumed that benefits are linear in the speed of signaling (that is, the inverse of the time to deliver the required number of signaling molecules). But there may be biological contexts in which benefits are saturating, strictly sub-linear, or even non-concave in the speed of signaling -- and this distinction could qualitatively change the performance of direct transport relative to mortal diffusion as a mechanism of signaling. A related set of questions pertain to a source cell that communicates with multiple target cells. In this setting, which is common in biological contexts, the benefits of successful signaling may again be sub- or super-additive across targets, depending upon whether reaching multiple targets is strictly required for producing a successful biological function, or merely additionally beneficial. Analysis of mean passage times and measures of efficiency in the setting of multiple targets remains an important and rich area for future research.

\section*{Acknowledgement}
We thank Tatyana Svitkina for her helpful comments. Y.M. was supported by the NSF (DMS-2042144) and the Simons Foundation (Math+X grant).

\bibliography{refs}
\end{document}